\documentclass[showpacs,preprint,amsmath,amssymb,aps]{revtex4}

\usepackage{graphicx}
\usepackage{dcolumn}
\usepackage{bm}

\begin{document}

\preprint{}
{\bf }\title{
Electromagnetic production of hypernuclei.
}
\author{B.~I.~S.~van der Ventel$^{1}$}
\author{T.~ Mart$^{2}$}
\author{H-F.~L\"{u}$^{3}$}
\author{H.~L.~Yadav$^{4}$}
\author {G.~C.~Hillhouse$^{1}$}  
\affiliation{$^{1}$Department of Physics, 
Stellenbosch University, Private Bag X1, Matieland 7602, South Africa}
\affiliation{$^{2}$Departemen Fisika, FMIPA, Universitas Indonesia, Depok,16424, Indonesia}
\affiliation{$^{3}$School of Science, Chinese Agriculture University, Beijing 100083}
\affiliation{$^{4}$Department of Physics, Banaras Hindu University, Varanasi
221005, India} 
\date{\today}

\begin{abstract} 
A formalism for the electromagnetic production of hypernuclei is developed where the cross section
is written as a contraction between a leptonic tensor and a hadronic tensor. The hadronic tensor is written in
a model-independent way by expanding it in terms of a set of five nuclear structure functions.
These structure functions are calculated by assuming that the virtual photon interacts with only
one bound nucleon. We use the most recent model for the elementary current operator which gives a good
description of the experimental data for the corresponding elementary process. The bound state
wave functions for the bound nucleon and hyperon are calculated within
a relativistic mean-field model. We calculate the unpolarized triple differential cross section for the hypernuclear 
production process
$ e + {^{12}{\rm C}} \, \,  \longrightarrow \, \,  e + K^{+} +  {^{12}_{~\Lambda}}{\rm B} $
as a function of the kaon scattering angle. The nuclear structure functions are calculated within a particle-hole
model. The cross section displays a characteristic form of being large for small values of the kaon 
scattering angle with a smooth fall-off to zero with increasing angle. The shape of the cross section is essentially
determined by the nuclear structure functions.
In addition, it is found that for the unpolarized triple differential cross section 
one structure function is negligible over the entire range of the kaon scattering angle.\\
Keywords: Hypernuclear; Relativistic Mean Field Models; Structure Functions; Strangeness Production
\end{abstract}

\pacs{24.10.Jv, 24.70.+s, 25.40.-h}

\maketitle

\section{\label{section_intro}Introduction}

The study of strange particles and hypernuclei remains an area of intense theoretical and experimental
activity. Hypernuclei represent an exotic state of matter since they contain particles with quantum numbers
such as strangeness which are not associated with ordinary nuclear matter. Whereas the nucleon-nucleon
interaction is very well-known from elastic scattering data, our knowledge of the hyperon-nucleon, and 
hyperon-hyperon interactions is still relatively incomplete. A major goal of nuclear physics should also be
a unified understanding of baryon-baryon interactions. Studies of hypernuclei are also important since they can
give us insight into the role of strangeness in the stellar environment. A free $\Lambda$ particle is unstable
and will primarily decay via the weak interaction to a nucleon-pion system. However, a $\Lambda$ in the nuclear
medium will interact strongly with the other nucleons, hence forming a hypernucleus. The $\Lambda$ is unaffected by
the Pauli exclusion principle and can therefore occupy any one of the states already filled by the nucleons. In a
one-boson-exchange picture the zero isospin of the $\Lambda$ forbids the exchange of isovector mesons such as a
pion or the rho meson with a nucleon, and therefore leads to a lack of strong tensor components in the 
$\Lambda N$ interaction. Consequently the $\Lambda N$ interaction is much weaker than the $NN$ interaction, and the
$\Lambda$ in the nucleus does not lead to a major disruption of the shell structure \cite{Gibson_PhysRep257_1995}.
Hypernuclei therefore provide remarkable experimental evidence for the shell-model of nuclear structure.

A wealth of experimental data on hypernuclei have been accumulated by making use of hadronic probes 
\cite{Gibson_PhysRep257_1995}.
These include the strangeness exchange reaction
\begin{eqnarray}
\label{eq_95}
 K^{-} \, (s \bar{u} \, ) + n \, (uud \, )
 & \longrightarrow &
 \Lambda \, (uds \, ) + \pi^{-} \, (d \bar{u} \, )
\end{eqnarray}
and the associate production process
\begin{eqnarray}
\label{eq_96}
 \pi^{+} \, (u \bar{d} \, ) + n \, (udd \, )
 & \longrightarrow &
 \Lambda \, (uds \, ) + K^{+} \, (u \bar{s} \, ).
\end{eqnarray}

An alternative production mechanism is through the use of real or virtual photons, i.e.,
\begin{eqnarray}
\label{eq_97}
 \gamma + p \, (uud \, )
 & \longrightarrow &
 K^{+} \, (u \bar{s} \, ) + \Lambda \, (uds \, ).
\end{eqnarray}
Hyperon production via the electromagnetic interaction requires the production of a strange quark/anti-quark pair. The
large momentum transfer for associate production decreases the sticking probability of the $\Lambda$, and consequently
the probability for obtaining a bound hypernuclear system in the final state \cite{Hinton_PhD_2000}. Since the
production of the reaction particles are limited to the very forward angles, it is necessary to detect the electron
and the kaon in coincidence \cite{Hungerford_NPA691_21_2001}.

However, electron beams offer a number of distinct advantages. Indeed,
with the advent of Jlab, our understanding of the role of the electromagnetic production process
has greatly increased. The precision of the electron beam, as well as good
spatial and energy resolution make up for the small $(e,e', K^{+})$ cross
section relative to hadronic production mechanisms \cite{Hinton_PhD_2000}.
The $(e,e', K^{+})$ reaction converts a proton in a target nucleus, and 
populates proton-hole $\Lambda$-particle states. This reaction therefore 
produces neutron-rich $\Lambda$ hypernuclei. Electroproduction excites both 
natural and unnatural parity states with comparable strength 
\cite{Cohen_PRC32_543_1985}. 
Both the photon and the $K^{+}$ interacts relatively weakly with the nucleus and
therefore the $(e,e' K^{+})$ reaction is not confined to the
nuclear surface, hence hypernuclear states can be studied with a 
deeply-bound hyperon. In heavier nuclei the behavior of a $\Lambda$ in
nuclear matter may be studied. The transition operator has a spin-part, hence
one can also probe spin-flip states. The electron beam can be polarized,
whereas no polarized $K^{-}$ (or $\pi^{+}$) beams exist 
\cite{Cohen_PRC32_543_1985}. 

The study of baryonic resonances is an important field in hadron phenomenology.
Theoretical work to determine the excitation spectrum of nucleons has been 
done mainly within the quark model framework. However, these models predict a
much richer spectrum than what has been observed with $\pi N \, \longrightarrow \,
\pi N$ scattering experiments. These missing resonances may therefore be identified by
studying the electromagnetic production of kaons and hyperons \cite{Mart_PRC61_012201_1999}.
In the electromagnetic production process, resonant baryon formation
and kaon exchange play a primary role. The coupling of $N^{*}$'s and 
$\Delta$'s to meson-hyperon final states may be studied, and compared to SU(3) flavor
symmetry predictions. In the case of electroproduction $(q^{2} \, \neq \, 0)$
two new features are introduced: (i) the longitudinal coupling of the photons
in the initial state, and (ii) the electromagnetic and hadronic form factors 
of the exchanged particles. 
In the case of electroproduction, the cross
sections for $\Lambda$ and $\Sigma^{0}$ production is quite different. This is
due to the isospin selectivity. In the $K^{+} \Lambda$ final states only the
$N^{*}$ resonances are allowed, whereas for $K^{+} \Sigma^{0}$ final states,
$\Delta$ resonances may also contribute to hyperon formation. 
More information about the elementary process
may be gleaned from electroproduction than from photoproduction, since the virtual
photon mass and polarization may be varied independently \cite{Niculescu_PRL81_1805_1998}.

A number of experiments have been performed over the years to investigate strangeness effects
in nuclear physics. See for example Table I in Ref.~\cite{Tamura:2004ci}. Indeed, the experimental
pursuit of the electromagnetic production of strangeness was given great impetus by the Jlab
facility \cite{Schumacher_NPA585_63_2001}. Longitudinal and transverse cross sections were 
measured for the reaction $^{1}H (e,e' K^{+}) \Lambda$ 
\cite{Niculescu_PRL81_1805_1998,Mohring_PRC67_055205_2003}. In experiment E89-009 the focus was 
on the production of hypernuclei \cite{Zhu_PhD_2001,Yuan_PhD_2002}. The hypernucleus 
$^{12}_{~\Lambda}$B was produced via the reaction $(e,e' K^{+})$ using high-energy electron beams 
\cite{Miyoshi_PRL90_232502_2003}. Experiment E91-016 focussed on
kaon electroproduction from deuterium \cite{Koltenuk_PhD_1999,Cha_PhD_2000}, as well as from
$^{3}$He and $^{4}$He targets \cite{Uzzle_PhD_2002,Dohrmann_PRL93_24501_2004}. The quasi-free
electroproduction of unbound $\Lambda$, $\Sigma^{0}$ and $\Sigma^{-}$ hyperons on
carbon and aluminum targets was studied in Ref.~\cite{Hinton_PhD_2000}. Strangeness production off
the proton and from nuclear targets has been investigated by the CLAS collaboration. Cross section and
recoil polarization data for the reactions $\gamma + p \, \longrightarrow \, K^{+} + \Lambda$ and
$\gamma + p \, \longrightarrow \, K^{+} + \Sigma^{0} $ for center-of-mass energies between 1.6 and 2.3 GeV
\cite{McNabb_PRC69_042201_2004}. 

The theoretical description of the elementary process is essential for studying hypernuclei
formation. Obtaining results directly from QCD is a formidable task, and the standard approach is to use
an effective field theory based on baryonic and mesonic degrees of freedom. The use of these so-called
isobaric models has been pursued by a number of authors 
\cite{Thom_PRC151_1322_1966,Deo_PRC9_288_1974,Cotanch_NPA450_419c_1986,Cotanch_PRC38_2691_1988,Tanabe_PRC39_741_1989,
Adelseck_PRC42_108_1990,Williams_PRC43_452_1991,Williams_PRC46_1617_1992,David_PRC53_2613_1995,Mizutani_PRC58_75_1997,
Bennhold:1999mt,Bydzovsky:2004dy,Mart:2004au,Mart_PRC61_012201_1999} with a summary of theoretical work given
in Ref.~\cite{Saghai:2003ta}. 

When making the transition from a theoretical description of the elementary process to that of hypernuclei
production, a number of additional complications enter. These include (i) the description of the current operator
in the nuclear medium, (ii) the nuclear structure model for the bound nucleons and hyperons, and (iii) the effect of 
nuclear distortion effects on the incoming and outgoing particles. The electromagnetic production of hypernuclei
has been investigated by a number of authors for photoproduction 
\cite{Hsiao_PRC28_1668_1983,Rosenthal:1987ju,Bennhold_PRC39_927_1989,Bennhold:1987nq,Lee_NPA695_237_1999,Motoba:1994sr,
Bennhold_PRC39_1994_1989,Bennhold_PRC43_775_1991} and electroproduction
\cite{Cohen_PRC32_543_1985,Sotona:1994st}. 
The first complication is addressed by invoking the impulse approximation, i.e., the elementary current operator is
assumed unchanged in the nuclear medium. The nuclear structure is described by non-relativistic shell model wave functions
or by solving the Dirac equation with scalar and vector potentials to obtain bound state nucleon and hyperon wave 
functions. Finally, the nuclear distortion effects are treated within an optical potential formalism.
In this work we adhere to the basic philosophy of these works, but also add a new feature, namely
we write the triple differential cross section as a contraction of a leptonic tensor and a hadronic tensor. Following
Refs.~\cite{Picklesimer_PRC34_1860_1986,VanderVentel_PRC69_035501_2004,VanderVentel_PRC73_025501_2006} we write
the hadronic tensor in terms of a set of five nuclear structure functions. Apart from the one-photon exchange
approximation, and the additional assumption that the virtual photon interacts with only one bound nucleon, the
formalism is still model-independent. In principle this allows one to study the cross section by doing a 
Rosenbluth-type separation. We do not explore this avenue in this work, but instead calculate the structure functions
using a definite model for the process occuring at the hadronic vertex.
Further, we calculate the bound state wave
functions for the nucleons and hyperons using three different relativistic mean-field models. These are the linear
Walecka model~\cite{SW86}, the successful NL3 parameter set~\cite{Lalazissis_PRC55_97}, and the recently-introduced FSUGold
parameter set~\cite{Tod05_PRLxx}. Finally, we employ the most up-to-date form of the elementary current operator that 
gives a satisfactory description of the elementary process \cite{Mart_PRC61_012201_1999}.
In this work we use the model for the corresponding elementary process, together
with the incorporation of nuclear structure effects, through an accurately-calibrated relativistic mean-field model. 
Thus, accurate binding energies and nucleon momentum distributions are employed. This method is based on the
impulse approximation and provides a fully relativistic study both in the reactive content and the nuclear structure. 
In addition, it provides a simple way to study medium modifications of the produced mesons, as well as the
resonances contributing to the elementary amplitude.

This paper is organized as follows. In Sec. \ref{section_formalism} we discuss the kinematics, and show how the
cross section may be written in terms of a contraction between the leptonic tensor and the hadronic tensor.
Most of the technical details are deferred to Appendices \ref{section_appendix1} and \ref{section_appendix2}.
The model for the elementary amplitude as well as the nuclear structure models are discussed in 
Secs.~\ref{section_elementary_amplitude} and \ref{section_nuclear_structure}. Results for both the free process
and the unpolarized triple differential cross section for hypernuclear production are given in
Sec.~\ref{section_results} with a summary in Sec.~\ref{section_summary}.

\section{\label{section_formalism}Formalism}

\subsection{\label{section_cross_section}Cross section and kinematics}

Consider the electromagnetic production of hypernuclei

\begin{figure}
\includegraphics[height=6cm,angle=0]{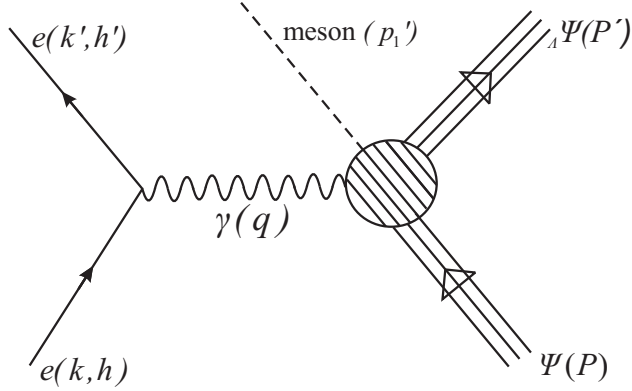}
 \caption{Lowest-order Feynman diagram for meson and hypernuclei
electromagnetic production.
\label{fig_1}}
\end{figure}

\begin{eqnarray}
\label{eq_1}
 e + {\rm A} & \longrightarrow & e + \mbox{meson} +  {_{\Lambda} {\rm B}}
\end{eqnarray}
which is shown schematically in Fig.~\ref{fig_1}. If we confine ourselves to the extreme
relativistic limit, then the incoming (outgoing) electrons may be specified by their
four-momenta and helicity, i.e., $(k,h)$ $[(k',h' \, )]$. In the one-photon exchange
approximation the reaction proceeds via the exchange of a virtual photon with
four-momentum $q^{\mu} \, = \, k^{\mu} - k'^{\mu} \, = \, (q_{0}, {\bf q} \, )$. The outgoing
meson is specified by its four-momentum $p_{1}'$. The target and residual hypernucleus have 
four-momenta $P$ and $P'$, respectively. The differential cross section may 
be written in terms
of these kinematical quantities and the transition matrix element ${\cal M}$ as
\begin{eqnarray}
\label{eq_2}
 d \sigma & = & \displaystyle \frac{1}{| {\bf v_{1}} - {\bf v_{2}} \, |} \,
\frac{d^{3} {\bf k}'}{(2 \pi)^{3}} \, \frac{d^{3} {\bf p_{1}'}}{2 E_{p_{1}'} (2 \pi)^{3}} \,
\frac{d^{3} {\bf P}'}{(2 \pi)^{3}} \, (2 \pi)^{4} \, \delta(k + P - k' - p_{1}' - P' \, ) \,
|{\cal M}|^{2}
\end{eqnarray}
where $| {\bf v_{1}} - {\bf v_{2}} \, |$ is the initial relative velocity. The transition 
amplitude ${\cal M}$ contains all the dynamical information about the reaction and will be studied in
detail in Secs.~\ref{section_tensors} and \ref{section_Wmunu_model}. In the laboratory frame the
initial flux for massless electrons is equal to one. The spatial part of the four-dimensional
Dirac delta function allows the integral over $d^{3} {\bf P'}$ to be performed. This fixes the
three-momentum of the residual hypernucleus to be 
\begin{eqnarray}
\label{eq_3}
 {\bf P'} & = & {\bf k} - {\bf k'} - {\bf p_{1}'} \, = \, {\bf q} - {\bf p_{1}'}
\end{eqnarray}
where ${\bf q} \, = \, {\bf k} - {\bf k'}$ is the three-momentum transfer to the nucleus. 
In Appendix~\ref{section_appendix1} it is shown that the triple differential cross 
section for the electromagnetic production of hypernuclei in the electron-nucleus laboratory frame is
given by
\begin{eqnarray}
\label{eq_16a}
 \displaystyle \frac{d \sigma}{d E_{k'} \, d \left( \cos \theta' \, \right) \,
 d \Omega_{1}'}
 & = &
 K \, |{\cal M} |^{2}
\end{eqnarray}
where $K$ is a kinematic quantity that is fully determined by the energies and masses of the 
reaction particles, as well as the scattering angles of the ejectiles [see Eq.~(\ref{eq_16})].

\subsection{\label{section_tensors}Triple differential cross section in terms of leptonic and hadronic
tensors}

Within the framework of the relativistic plane wave impulse approximation,
the transition matrix element ${\cal M}$ for the electromagnetic production of hypernuclei
may be defined as
\begin{eqnarray}
\label{eq_19a}
 {\cal M} & = & \left[ \overline{U} ({\bf k'}, h' \, ) \gamma_{\mu} \, 
 U ({\bf k},h) \, \right] \, \left( \frac{e^{2}}{q^{2}} \right) \,
 \langle \, p_{1}'; \, _{\Lambda} \hspace{-0.10cm} \Psi (P' \, ) \, | \hat{J}^{\mu} (q) \, | 
 \Psi(P) \, \rangle
\end{eqnarray}
with $e^{2}/4 \pi \, = \, 1/137$. 
In Eq. (\ref{eq_19a}) $\hat{J}^{\mu}$ is the nuclear current operator, and 
$U({\bf k},h)$ is the plane wave Dirac spinor (defined in Eq.~(\ref{eq_20}))
for the incident or ejectile electrons. $| \Psi(P) \, \rangle$ represents the many-body
state for the incident nucleus, and $| p_{1}'; \Psi(P' \, ) \, \rangle$ represents the
final state consisting of the many-body residual hypernucleus state, and the outgoing
meson. Using Eq.~(\ref{eq_19a}) it follows that
\begin{eqnarray}
\label{eq_22a}
 |{\cal M}|^{2} \, = \, {\cal M} \, {\cal M}^{*} 
 & = &
 \left( \frac{e^{2}}{q^{2}} \right)^{2} \, \ell_{\mu \nu} \, {\cal W}^{\mu \nu}
\end{eqnarray}
where we have introduced the leptonic tensor
\begin{eqnarray}
\label{eq_23a}
 \ell_{\mu \nu}
 & = & 
 \left[ \, \overline{U} ({\bf k'}, h' \, ) \gamma_{\mu} \, U ({\bf k},h) \, \right] \,
 \left[ \, \overline{U} ({\bf k'}, h' \, ) \gamma_{\nu} \, U ({\bf k},h) \, \right]^{*}
\end{eqnarray}
and the hadronic tensor
\begin{eqnarray}
\label{eq_24a}
 {\cal W}^{\mu \nu}
 & = &
 \langle \, p_{1}'; \, _{\Lambda} \hspace{-0.10cm} \Psi (P' \, ) \, | \hat{J}^{\mu} (q) \, | 
 \Psi(P) \, \rangle \,
 \langle \, p_{1}'; \, _{\Lambda} \hspace{-0.10cm} \Psi (P' \, ) \, | \hat{J}^{\nu} (q) \, | 
 \Psi(P) \, \rangle^{*}.
\end{eqnarray}
These two tensors are studied in detail in Appendix \ref{section_appendix2}.
In addition, it is shown in Appendix \ref{section_appendix2} that the unpolarized 
triple differential cross section for electromagnetic hypernuclei production in the
electron-nucleus laboratory frame is given by
\begin{eqnarray}
\nonumber
 \displaystyle \frac{d \sigma}{d E_{k'} \, d \left( \cos \theta' \, \right) \,
 d \Omega_{1}'}
 & = &
 K \,
 \left( \frac{e^{2}}{q^{2}} \right)^{2} \, \ell_{\mu \nu}^{(0)} \, {\cal W}^{\mu \nu}_{S}
\\
\nonumber
 & = & K \, 
 \displaystyle \left( \frac{e^{2}}{q^{2}} \right)^{2} \, \frac{1}{E_{k} \, E_{k'}} \,
 \left[ W_{1} \, \left( -3 k \cdot k' + 2 f_{1} (k,k' \, ) \, \right) +
 W_{2} \, \left( -k \cdot k' \, f_{1} (P,P) + \right. \right.
\\
\nonumber
 & &
 \left. 2 f_{1} (k,P) \, f_{1} (k',P) \, \right) + W_{3} \, 
 \left( -k \cdot k' \, f_{1} (p_{1}', p_{1}' \, ) + 2 f_{1} (k, p_{1}') \, f_{1} (k', p_{1}' \, ) \right)  +
\\
\label{eq_16b}
 & &
\left. W_{4} \, \left( 2 f_{2} (P,p_{1}' \, ) \, \right) \, 
\right]
\end{eqnarray}
where $K$ is a kinematic quantity that is fully determined by the energies and masses of the 
reaction particles, as well as the scattering angles of the ejectiles (see Eq.~(\ref{eq_16})).
The functions $f_{i}$ are defined in Eq.~(\ref{eq_52}) and (\ref{eq_53}).
Apart from the one-photon exchange approximation and the additional assumption that the virtual
photon interacts with only one bound nucleon, Eq.~(\ref{eq_16b}) is still model-independent.
It shows that the unpolarized triple differential cross section may be determined from purely
kinematical quantities, and a set of four nuclear structure functions, $W_{1}$ to $W_{4}$. 
In this sense 
Eq.~(\ref{eq_16b}) is in line with the philosophy of Refs.~\cite{Picklesimer_PRC32_1312_1985,
VanderVentel_PRC69_035501_2004,VanderVentel_PRC73_025501_2006}. 
In the following
section we present a calculation of these structure functions by evaluating the matrix element
\begin{eqnarray}
\label{eq_16c}
 \langle \, p_{1}'; \, _{\Lambda} \hspace{-0.10cm} \Psi (P' \, ) \, | \hat{J}^{\mu} (q) \, | 
 \Psi(P) \, \rangle
\end{eqnarray}
in a model-dependent way.

\subsection{\label{section_Wmunu_model}Model-dependent form of the hadronic tensor}

In the previous section a general formalism was developed for the electromagnetic
production of hypernuclei. We now present a model-dependent evaluation of
the structure functions $W_{i}$.

The exact expression for the hadronic tensor is given in Eq. (\ref{eq_24}), and is
defined in terms of the following matrix element (and its complex conjugate):
\begin{eqnarray}
\label{eq_54}
 J^{\mu} & = &
 \langle \, p_{1}'; \, _{\Lambda} \hspace{-0.10cm} \Psi (P' \, ) \, | \hat{J}^{\mu} (q) \, | 
 \Psi(P) \, \rangle.
\end{eqnarray}
\begin{figure}
\includegraphics[height=6cm,angle=0]{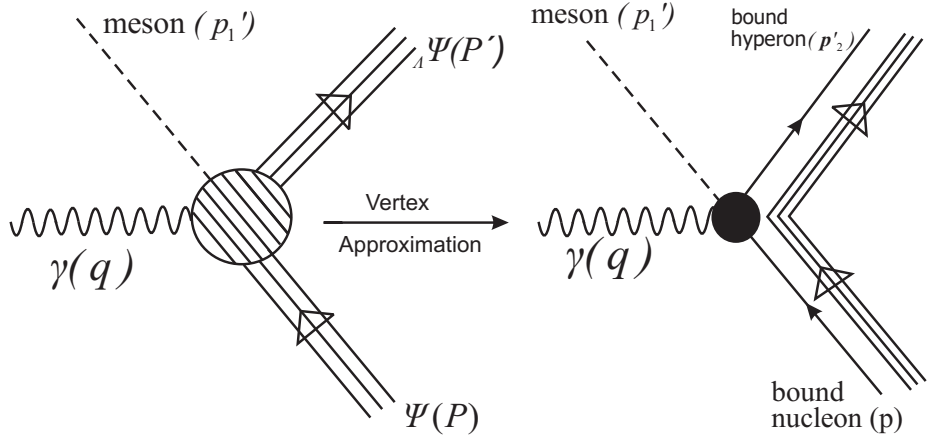}
 \caption{Graphical representation for the approximation employed at the 
hadronic vertex in order to obtain a tractable form for the matrix element
given in Eq. (\ref{eq_16c}).
\label{fig_3}}
\end{figure}

To obtain a tractable form for this extremely complicated object, we rely on a number of
approximations which are depicted schematically in Fig.~\ref{fig_3}. The principle
assumption is that the virtual photon interacts with only one bound nucleon. This
neglects two- and many-body components of the electromagnetic current operator. Secondly,
it is assumed that the resulting meson and hyperons are produced from the interaction between
the virtual photon and the nucleon to which it had coupled. This neglects two- and many-body
rescattering processes. Additionally, nuclear distortion effects on the kaon are
neglected. These simplifying assumptions lead to the following expression for the
hadronic matrix element
\begin{eqnarray}
\label{eq_55}
 J^{\mu}_{\alpha_{\Lambda}, \alpha_{N}}
 & = &
 \int \, d^{3} {\bf p''} \, d^{3} {\bf p} \, \delta ({\bf q} + {\bf p} - {\bf p_{1}'} -
 {\bf p''}) \, \overline{{\cal U}}_{\alpha_{\Lambda}} ({\bf p''})
 \, \hat{J}^{\mu} (q) \, {\cal U}_{\alpha_{N}} ({\bf p})
\\
\label{eq_56}
 & = & 
 \int \, d^{3} \, {\bf p} \, \overline{{\cal U}}_{\alpha_{\Lambda}} 
 ({\bf q } + {\bf p} - {\bf p_{1}'}) \, \hat{J}^{\mu} (q) \, 
 {\cal U}_{\alpha_{N}} ({\bf p})
\\
\label{eq_57}
 & = &
 \int_{0}^{1} dp'' \int_{0}^{1} d \theta'' \int_{0}^{1} d \phi''
 \left( 2 \pi^{2} p_{\rm max} \right) \, \left( p^{2} \sin \theta \right) \,
 \overline{{\cal U}}_{\alpha_{\Lambda}} ({\bf q} + {\bf p} - {\bf p_{1}'}) \,
 \hat{J}^{\mu} (q) \, {\cal U}_{\alpha_{N}} ({\bf p})
\end{eqnarray}
where 
\begin{eqnarray}
\label{eq_58}
 p & = & p_{\rm max} p'',
\end{eqnarray}
\begin{eqnarray}
\label{eq_59}
 \theta & = & \pi \, \theta'',
\end{eqnarray}
\begin{eqnarray}
\label{eq_60}
 \phi & = & 2 \pi \, \phi''.
\end{eqnarray}
In Eqs. (\ref{eq_55}) - (\ref{eq_57}) the labels $\alpha_{\Lambda}$ and $\alpha_{N}$ refer to
the quantum numbers necessary to specify the bound state wave functions of the nucleon and
hyperon, respectively. 
In Eq. (\ref{eq_58}) $p_{\rm max}$ refers to the maximum momentum for which the momentum space
wave function is still appreciable. More detail will be provided in Sec.~\ref{section_nuclear_structure}.
The operator $\hat{J}^{\mu}$ refers to the current operator for the corresponding elementary
process. The model for $J^{\mu}$ given in Eq.~(\ref{eq_57}) limits our results to simple particle-hole
configurations of the produced hypernucleus. The model-dependent hadronic tensor is then given by
\begin{eqnarray}
\label{eq_60a}
 {\cal W}^{\mu \nu}_{model}
 & = &
 \sum_{\alpha_{N}, \alpha_{\Lambda}} \, J^{\mu}_{\alpha_{N}, \alpha_{\Lambda}} \,
 \left( J^{\nu}_{\alpha_{N}, \alpha_{\Lambda}} \right)^{*}.
\end{eqnarray}
The model-dependent structure functions are then determined from the equation below 
(see also Eq.~(\ref{eq_44})):
\begin{eqnarray}
\label{eq_60b}
 \underline{W} \, = \,
 \left(
  \begin{array}{c}
   W_{1}
   \\[0.25mm]
   W_{2}
   \\[0.25mm]
   W_{3}
   \\[0.25mm]
   W_{4}
  \end{array}
 \right)
 & = &
 \left( U^{-1} \right) \, u_{i, \mu \nu} \, {\cal W}^{\mu \nu}_{model}
\end{eqnarray}
where the $4 \times 4$ matrix $U$ is given by Eq.~(\ref{eq_46}).
The two critical ingredients namely the bound state wave functions and the elementary
current operator will be discussed in the following two sections.

\subsubsection{\label{section_nuclear_structure}Nuclear structure}

In this work the bound state wave functions for the nucleon and the lambda are both determined
from relativistic mean-field theory. For the nucleon bound state wave functions we use the linear
Walecka model~\cite{SW86}, the successful NL3 parameter set~\cite{Lalazissis_PRC55_97}, and the 
recently-introduced FSUGold parameter set~\cite{Tod05_PRLxx}. For the hyperon state we employed the
Lagrangian density of Ref.~\cite{Lu03}.

For spherically symmetric nuclei the single-particle bound state
wave function in position space is given by
\begin{eqnarray}
\label{eq_61}
 {\cal U}_{\alpha} ({\bf x}) & = & {\cal U}_{E \kappa m} ({\bf x}) \, = \,
 \left(
  \begin{array}{c}
   \left[ \displaystyle \frac{g_{E \kappa} (r)}{r} \right] \, {\cal Y}_{+\kappa m} ({\bf \hat{x}})
   \\[0.5cm]
   \left[ \displaystyle \frac{i f_{E \kappa} (r)}{r} \right] \, {\cal Y}_{-\kappa m} ({\bf \hat{x}})
  \end{array}
 \right)
\end{eqnarray}
where $m$ is the magnetic quantum number, $E$ is the binding energy, $\kappa$ the generalized
angular momentum, and the spinor-spherical harmonics are defined as
\begin{eqnarray}
\label{eq_62}
 {\cal Y}_{\kappa m} ({\bf \hat{x}})
 & = &
 \sum_{s_{z}' \, = \, \pm \, 1/2} \, \langle l \frac{1}{2}, m - s_{z}', s_{z}', | j m \, \rangle \,
 Y_{l, m - s_{z}'} ({\bf \hat{x}}) \, \chi_{s_{z}'}.
\end{eqnarray}
The orbital angular momentum $l$, and the total angular momentum $j$, may be obtained as follows
\begin{eqnarray}
\label{eq_63}
 j & = & |\kappa| - \displaystyle \frac{1}{2}
\end{eqnarray}
and
\begin{eqnarray}
\label{eq_64}
 l & = & 
 \left \{
  \begin{array}{c}
   \kappa, \qquad \kappa \, > \, 0
   \\[0.25cm]
   -1 - \kappa, \qquad \kappa \, < \, 0.
  \end{array}
 \right.
\end{eqnarray}
The momentum space bound state wave function is defined by
\begin{eqnarray}
\label{eq_65}
 {\cal U}_{E \kappa m} ({\bf p})
 & = &
 \int \, d^{3} {\bf p} \, e^{-i {\bf p} \cdot {\bf x}} \,
 {\cal U}_{E \kappa m} ({\bf x})
\\
\label{eq_66}
 & = &
 4 \pi \left( -i \right)^{l} \,
 \left(
  \begin{array}{c}
   g_{E \kappa} (p) \, {\cal Y}_{\kappa m} ({\bf \hat{p}})
   \\[0.25cm]
   f_{E \kappa} (p) \, {\bf \sigma} \cdot {\bf \hat{p}} \, {\cal Y}_{\kappa m} ({\bf \hat{p}})
  \end{array}
 \right)
\end{eqnarray}
where
\begin{eqnarray}
\label{eq_67}
 g_{E \kappa} (p) 
 & = &
 \int_{0}^{\infty} \, dr \, \, r g_{E \kappa} (r) \, j_{l} (pr)
\end{eqnarray}
and
\begin{eqnarray}
\label{eq_68}
 f_{E \kappa} (p)
 & = &
 \mbox{sgn} (\kappa) \, \int_{0}^{\infty} \, dr \, \, r f_{E \kappa} (r) \,
 j_{2j -l} (pr)
\end{eqnarray}
where $j_{l} (z)$ is the spherical Bessel function.

\subsubsection{\label{section_elementary_amplitude}Elementary scattering operator}

The approximation depicted in Fig.~\ref{fig_3} shows that the electromagnetic
hypernuclei production process is essentially determined by the elementary
process
\begin{eqnarray}
\label{eq_70}
 \gamma \, \mbox{(virtual)} \, + \, \mbox{nucleon}
 & \longrightarrow &
 \mbox{meson} \, + \, \mbox{hyperon}.
\end{eqnarray}
In this work we invoke the impulse approximation and employ the model for 
$\hat{J}^{\mu}$ as discussed in Ref.~\cite{Mart_PRC61_012201_1999}. 
There are six different reaction
channels which may be explored using this formalism, namely
\begin{eqnarray}
\label{eq_71}
 e + p & \longrightarrow & e + K^{+} + \Lambda
\\
\label{eq_72}
 e + n & \longrightarrow & e + K^{0} + \Lambda
\\
\label{eq_73}
 e + p & \longrightarrow & e + K^{+} + \Sigma^{0}
\\
\label{eq_74}
 e + p & \longrightarrow & e + K^{0} + \Sigma^{+}
\\
\label{eq_75}
 e + n & \longrightarrow & e + K^{+} + \Sigma^{-}
\\
\label{eq_76}
 e + n & \longrightarrow & e + K^{0} + \Sigma^{0}.
\end{eqnarray}
An application of the formalism in this paper will however, only be to the production
of $\Lambda$ hypernuclei.
The electromagnetic current operator for each of these reaction channels
is written as 
\begin{eqnarray}
\label{eq_77}
 \hat{J}^{\mu}
 & = &
 \sum_{i \, = \, 1}^{6} \, A_{i} (s,t,q^{2} \, ) \, M_{i}^{\mu}
\end{eqnarray}
where
\begin{eqnarray}
\label{eq_78}
 M_{1}^{\mu} & = & \displaystyle \frac{1}{2} \, \gamma^{5} \, m_{1}^{\mu} \, = \,
 \displaystyle \frac{1}{2} \gamma^{5} \, 
 \left( \gamma^{\mu} \rlap /q - \rlap /q \gamma^{\mu} \right)
\\
\label{eq_79}
 M_{2}^{\mu} & = & \gamma^{5} \, m_{2}^{\mu} \, = \,
 \gamma^{5} \, \left[ \displaystyle \frac{1}{2} \left( p \cdot q + p_{2}' \cdot q \right) 
 \left( 2 p_{1}'^{\mu} - q^{\mu} \, \right) -\displaystyle \frac{1}{2} 
 \left( 2 p_{1}' \cdot q - q^{2} \right) \, \left( p^{\mu} + p_{2}'^{\mu} \, \right) \, \right]
\\
\label{eq_80}
 M_{3}^{\mu} & = & \gamma^{5} \, m_{3}^{\mu} \, = \, 
 \gamma^{5} \, \left( p_{1}' \cdot q \, \gamma^{\mu} - p_{1}'^{\mu} \, \rlap /q \right)
\\
\label{eq_81}
 M_{4}^{\mu} & = & -i \epsilon_{\alpha \lambda \beta \nu} \, p_{1}'^{\beta} \, q^{\nu} \,
 \gamma^{\alpha} \, g^{\mu \lambda}
\\
\label{eq_82}
 M_{5}^{\mu} & = & \gamma^{5} \, m_{5}^{\mu} \, = \, 
 \gamma^{5} \, \left( p_{1}'^{\mu} q^{2} - p_{1}' \cdot q \, q^{\mu} \, \right)
\\
\label{eq_83}
 M_{6}^{\mu} & = & \gamma^{5} \, m_{6}^{\mu} \, = \, \gamma^{5} \,
 \left( q^{\mu} \rlap /q - q^{2} \gamma^{\mu} \, \right).
\end{eqnarray}
In Eq. (\ref{eq_77}) $s$ and $t$ are the usual Mandelstam variables defined as
\begin{eqnarray}
\label{eq_84}
 s & = & \left( q + p \right)^{2} \, = \, \left( p_{1}' + p_{2}' \, \right)^{2}
\\
\label{eq_85}
 t & = & \left( q - p_{1}' \, \right)^{2} \, = \, \left( p_{2}' - p \right)^{2}
\\
\label{eq_86}
 u & = & \left( q - p_{2}' \, \right)^{2} \, = \, \left( p_{1}' - p \right)^{2}.
\end{eqnarray}
In Eqs.~(\ref{eq_84}) to (\ref{eq_86}) the four-momenta of the bound nucleon and bound
hyperon are denoted by $p$ and $p_{2}'$, respectively. The four-momentum of the bound
nucleon is defined as
\begin{eqnarray}
\label{eq_87}
 p^{\mu} & = & \left( M_{p} - E_{B}^{(N)}, {\bf p} \right) \, = \,
 \left( M_{p} - E_{B}^{(N)}, p \sin \theta \cos \phi, p \sin \theta \sin \phi,
 p \cos \theta \, \right)
\end{eqnarray}
where $p$, $\theta$ and $\phi$ refer to the integration variables defined in Eq. (\ref{eq_57}).
The invariant amplitudes $A_{i}$ are determined using an isobar model 
\cite{Mart_PRC61_012201_1999}. Feynman diagrams are written down for the $s$-, $t$- and
$u$-channels for kaon electroproduction from the nucleon. The following resonances are included
$S_{11}(1650), P_{11}(1710), P_{13}(1720), D_{13}(1895), K^{*}(892)$ and $K_{1}(1270)$.

\section{\label{section_results}Results}

As a first application of the formalism developed in
Sec.~\ref{section_formalism} we calculate the unpolarized triple
differential cross section for the hypernuclear production process
\begin{eqnarray}
\label{eq_88}
 e + {^{12}{\rm C}} & \longrightarrow & e + K^{+} + {^{12}_{~\Lambda}{\rm B}}.
\end{eqnarray}
This was the first hypernuclear spectroscopy experiment via
electroproduction performed at Jlab \cite{Miyoshi_PRL90_232502_2003}. 
Before presenting cross section results we first need to investigate
the two critical components that enter our formalism: (i) the
elementary operator $\hat{J}^{\mu}$, and (ii) the nuclear structure
input, namely the bound state wave functions for the bound nucleon and
bound hyperon.

\begin{figure}
\includegraphics[height=10cm,angle=0]{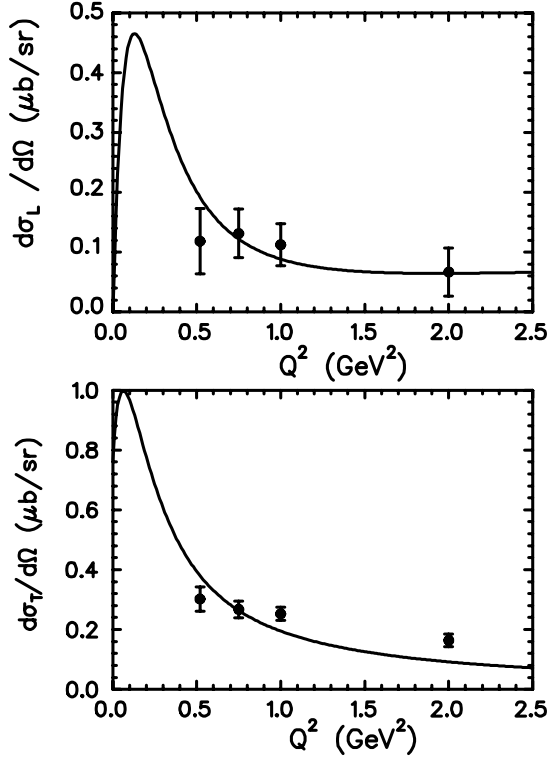}
 \caption{Longitudinal (top figure) and unpolarized transverse (bottom figure) differential cross sections
for the reaction $e + p \, \longrightarrow \, e + K^{+} + \Lambda$, as a function of $Q^{2} \, = \, -q^{2}$
(four-momentum transfer). The kinematical quantities are $W \, = \, \sqrt{s} \, = \, 1.84$ GeV and the kaon
scattering angle $\theta_{1}' \, = \, 0^{\circ}$. The data are from Ref. \cite{Mohring_PRC67_055205_2003}.
\label{fig_4}}
\end{figure}

The underlying elementary process for reaction (\ref{eq_88}) is
\begin{eqnarray}
\label{eq_89}
 e + p & \longrightarrow & e + K^{+} + \Lambda.
\end{eqnarray}
 It can be shown that the triple differential
cross section for the elementary hyperon production process can be written as 
\cite{Williams_PRC46_1617_1992}:
\begin{eqnarray}
\label{eq_90}
 \displaystyle \frac{d \sigma}{d E_{k'} \, d (\cos \theta' \, ) \, d \Omega_{1}'}
 & = &
 K_{lep} \, \displaystyle \frac{d \sigma_{V}}{d \Omega_{1}'}
\end{eqnarray}
where $K_{lep}$ is a kinematic factor that is purely described by the leptonic kinematics,
and $d \sigma_{V}/ d \Omega_{1}'$ represents the differential cross section for kaon production
from a virtual photon. This can be expanded in four terms which are related to the polarization of
the virtual photon. In particular for longitudinally polarized virtual photons we have that 
$d \sigma_{L}/ d \Omega_{1}'$ may be written in terms of the $\mu \, = \, 3$ and 
$\nu \, = \, 3$ components of the hadronic tensor \cite{Williams_PRC46_1617_1992}, i.e.,
\begin{eqnarray}
\label{eq_91}
 \displaystyle \frac{d \sigma_{L}}{d \Omega_{1}'}
 & = &
 \left[ \displaystyle \frac{-2 |{\bf p_{1}'} \, | \, \sqrt{s} \, q^{2}}
 {\left( s - M^{2} \, \right) \, q_{0}^{2}} \right] \,
 \left[ \displaystyle \frac{E_{p_{1}} \, E_{p_{2}'}}{2s} \right] \, {\cal W}^{33}.
\end{eqnarray}
Similarly the unpolarized transverse cross section is given by
\begin{eqnarray}
\label{eq_91a}
 \displaystyle \frac{d \sigma_{T}}{d \Omega_{1}'}
 & = &
 \displaystyle \frac{|{\bf p_{1}'} \, | \sqrt{s} \,}{2 (s - M^{2})} \,
 \left( {\cal W}^{11} + {\cal W}^{22} \right). 
\end{eqnarray}
In Eqs.~(\ref{eq_91}) and (\ref{eq_91a})
the hadronic tensor ${\cal W}^{\mu \nu}$ for the elementary process is defined as
\begin{eqnarray}
\label{eq_92}
 {\cal W}^{\mu \nu}
 & = &
 \displaystyle \sum_{s_{1}, s_{2}'} \, \hat{J}^{\mu} \, \left( \hat{J}^{\nu} \, \right)^{*}
\\
 & = &
\nonumber
 \displaystyle \sum_{s_{1}, s_{2}'} \, \left \{ \overline{U} ({\bf p_{2}'}, s_{2}' \, ) \,
 \left( \sum_{i=1}^{6} \, A_{i} \, M_{i}^{\mu} \, \right) \, U ({\bf p_{1}, s_{1}}) \, \right \}
\\
 & &
\label{eq_93}
 \displaystyle \left \{ \overline{U} ({\bf p_{2}'}, s_{2}' \, ) \,
 \left( \sum_{j=1}^{6} \, A_{j} \, M_{j}^{\nu} \, \right) \, U ({\bf p_{1}, s_{1}}) \, \right \}^{*}
\\
 & = &
\label{eq_94}
 \displaystyle \sum_{i,j=1}^{6} \, A_{i} \, A_{j}^{*} \, 
 \mbox{Tr} \, \left[ M_{i}^{\mu} \, \left( \rlap/ p_{1} + M \right) \, 
 \overline{M_{j}^{\nu}} \, \left( \rlap/{p'}_{2} + M_{Y} \right) \, \right].
\end{eqnarray}
In Eqs.~(\ref{eq_92}) - (\ref{eq_94}) we have summed over the spin projections of the
initial proton and the outgoing hyperon. For the elementary process the baryons may be
represented by plane wave Dirac spinors $U({\bf p}, s)$. The hadronic tensor was calculated 
in two independent ways: (i) from Eq.~(\ref{eq_93}) by explicitly programming the Dirac
spinors $U({\bf p}, s)$ , the matrices $M_{i}^{\mu}$ (defined in Eqs.~(\ref{eq_78}) to 
(\ref{eq_83})) and the current operator $\hat{J}^{\mu}$ (defined in Eq.~(\ref{eq_77})), and
(ii) from Eq.~(\ref{eq_94}) by explicitly performing the trace algebra over the
Dirac matrices (i.e., the matrices $M_{i}^{\mu}$). Both methods give identical results.
This ensures that the current operator $\hat{J}^{\mu}$ inserted in Eq.~(\ref{eq_57}) has been
correctly implemented numerically. In Fig.~\ref{fig_4} we show a graph of the longitudinal
and unpolarized transverse cross sections as a function of $Q^{2} \, = \, -q^{2}$ for the elementary
process $e + p \, \longrightarrow \, e + K^{+} + \Lambda$ with $W \, = \, \sqrt{s} \, = \, 1.84$ GeV
and the kaon scattering angle $\theta_{1}' \, = \, 0^{\circ}$. The data are from 
Ref.~\cite{Mohring_PRC67_055205_2003}.
The relatively good prediction of the scarce data by our model for the
elementary current operator provides the confidence to embed this
operator in the nuclear medium, for describing reactions on nuclei, and
therefore obtain quantitative results for the triple differential cross section for hypernuclei
production.

\begin{figure}
\includegraphics[height=10cm,angle=0]{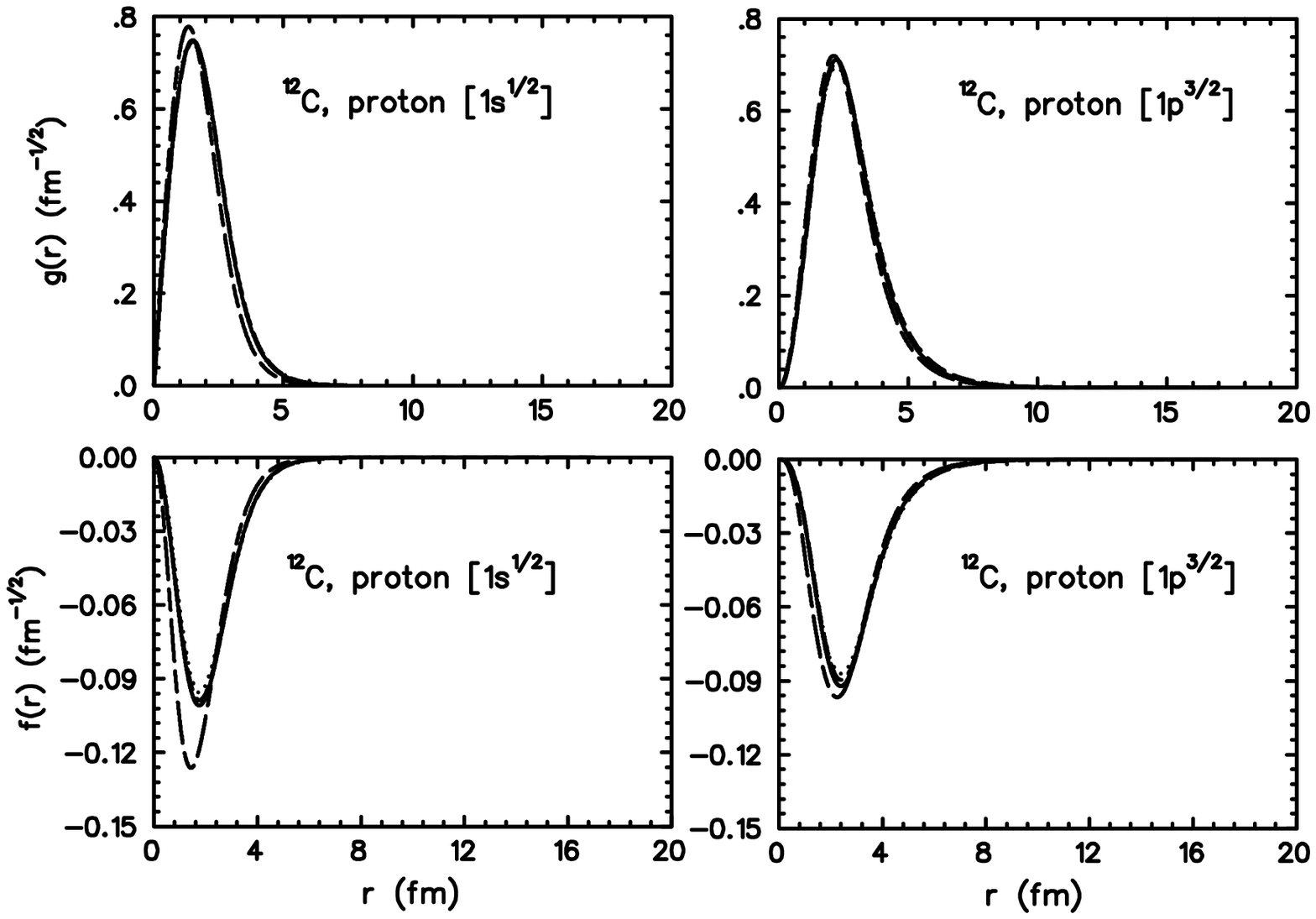}
 \caption{Upper $(g(r))$ and lower $(f(r))$ radial wave functions in position
space for the $1s^{1/2}$ and $1p^{3/2}$ proton orbitals of $^{12}$C. The solid
and dashed lines represent linear Walecka model predictions \cite{SW86}, the
long-dashed--short-dashed represents the NL3 calculation \cite{Lalazissis_PRC55_97},
and the dotted line represents the FSUGold model prediction \cite{Tod05_PRLxx}.
\label{fig_5}}
\end{figure}

\begin{figure}
\includegraphics[height=10cm,angle=0]{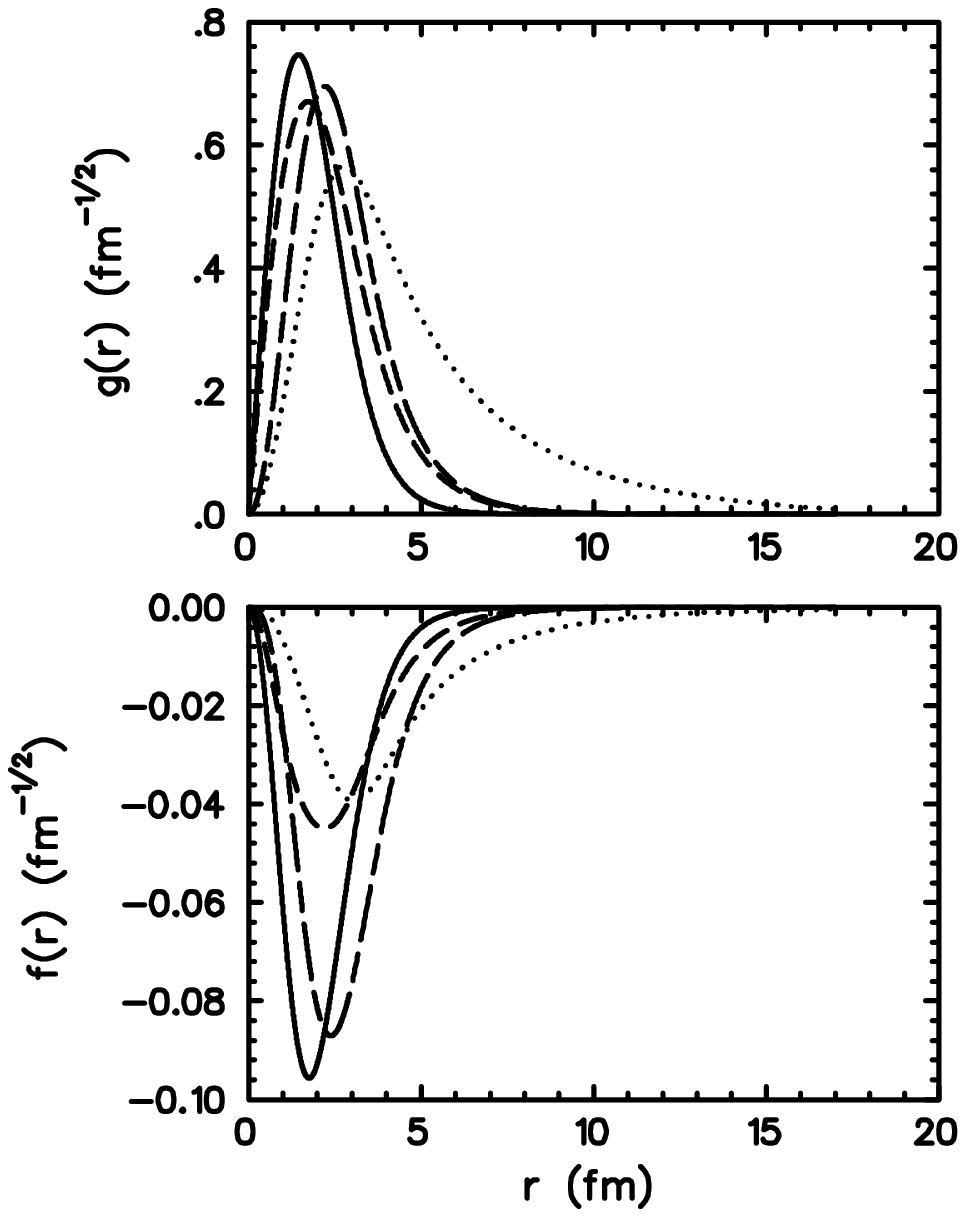}
 \caption{Upper $(g(r))$ and lower $(f(r))$ radial wave functions in position
space for the $1s^{1/2}$ and $1p^{3/2}$ orbitals of $^{12}$C and
\hspace{0.15cm} $_{\Lambda}^{12}$ \hspace{-0.25cm} B. The solid line
represents the $1s^{1/2}$ proton orbital of $^{12}$C, the dashed line the
$1s^{1/2}$ orbital of \hspace{0.15cm} $_{\Lambda}^{12}$ \hspace{-0.25cm} B,
the long-dashed--short-dashed line represents the $1p^{3/2}$ orbital of
$^{12}$C, and the dotted line the $1p^{3/2}$ orbital of
\hspace{0.15cm} $_{\Lambda}^{12}$ \hspace{-0.25cm} B. For this figure we only
employed the FSUGold model.
\label{fig_6}}
\end{figure}

\begin{figure}
\includegraphics[height=10cm,angle=0]{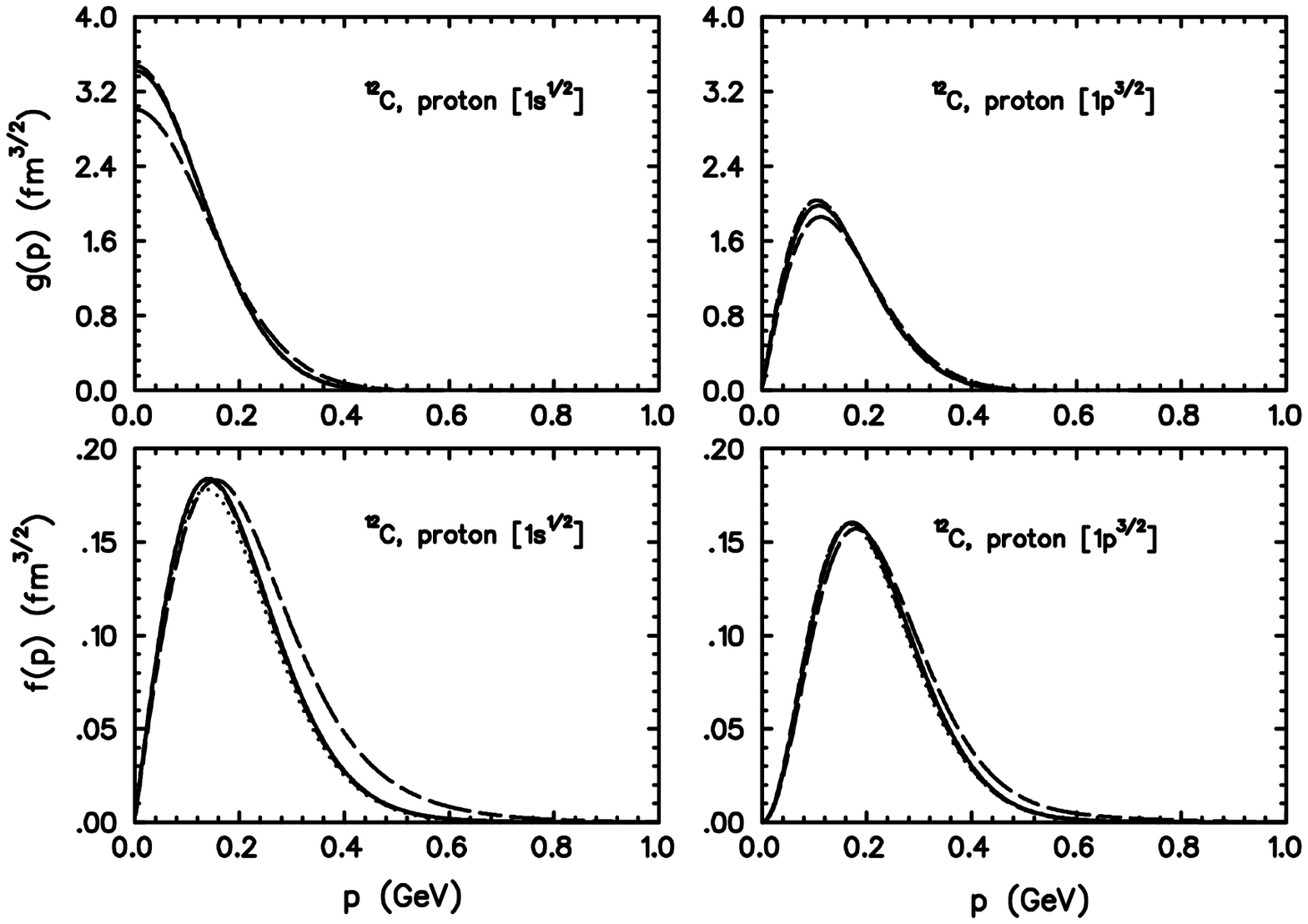}
 \caption{Upper $(g(p))$ and lower $(f(p))$ radial wave functions in momentum
space for the $1s^{1/2}$ and $1p^{3/2}$ proton orbitals of $^{12}$C. The solid
and dashed lines represent linear Walecka model predictions \cite{SW86}, the
long-dashed--short-dashed represents the NL3 calculation \cite{Lalazissis_PRC55_97},
and the dotted line represents the FSUGold model prediction \cite{Tod05_PRLxx}.
\label{fig_7}}
\end{figure}

In our formalism nuclear structure effects enter exclusively in terms
of the momentum distribution of the bound nucleons and hyperons, and
are calculated within a relativistic mean-field approximation. 
As was mentioned previously, there were three
models that we considered, namely the linear Walecka model~\cite{SW86}, the successful NL3 parameter
set~\cite{Lalazissis_PRC55_97}, and the recently-introduced FSUGold
parameter set~\cite{Tod05_PRLxx}. 
In Fig.~\ref{fig_5} we show the results for the upper $(g(r))$ and lower $(f(r))$ radial wave functions in 
position space
obtained from these three models for the $1s^{1/2}$ and $1p^{3/2}$ proton orbitals of $^{12}$C.
At the wave function level there is no real discernable difference between the different model
predictions. The upper $(g(r))$ and
lower $(f(r))$ proton wave functions (employing only the FSUGold
model) together with the upper and lower lambda wave functions are displayed in
Fig.~\ref{fig_6}. 
The momentum space wave functions are calculated from
Eqs.~(\ref{eq_67}) and (\ref{eq_68}). The results are shown in Fig.~\ref{fig_7} for the upper $g(p)$
and lower $(f(p))$ momentum space wave functions. Once again there is very little difference between
the models. We note that the wave functions are appreciable only for $p \, \approx \, 0.6$ GeV. This
fixes the parameter $p_{\rm max}$ referred to in Eqs.~(\ref{eq_57}) and (\ref{eq_58}). 

\begin{figure}
\includegraphics[height=10cm,angle=0]{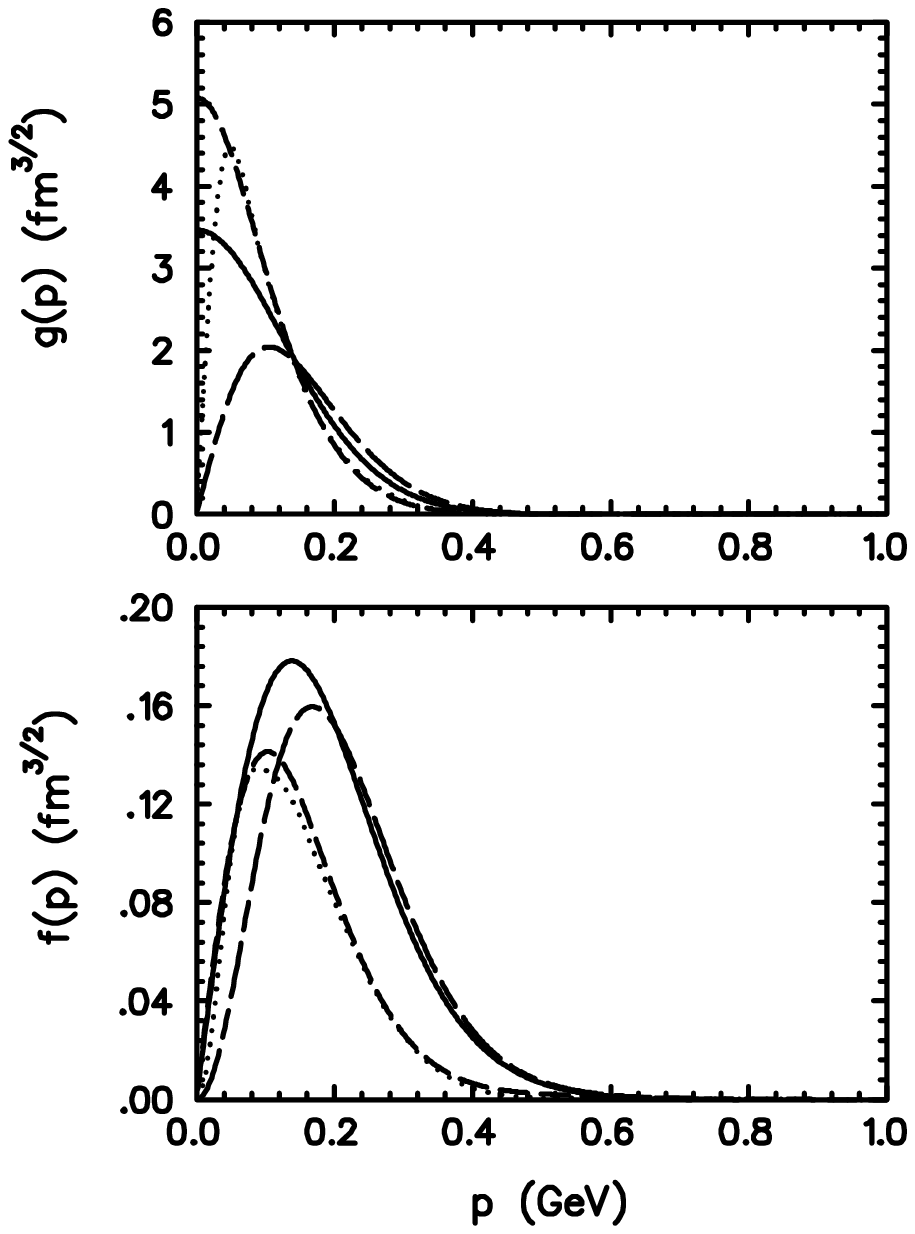}
 \caption{Upper $(g(p))$ and lower $(f(p))$ radial wave functions in momentum
space for the $1s^{1/2}$ and $1p^{3/2}$ orbitals of $^{12}$C and
\hspace{0.15cm} $_{\Lambda}^{12}$ \hspace{-0.25cm} B. The solid line
represents the $1s^{1/2}$ orbital of $^{12}$C, the dashed line the
$1s^{1/2}$ orbital of \hspace{0.15cm} $_{\Lambda}^{12}$ \hspace{-0.25cm} B,
the long-dashed--short-dashed line represents the $1p^{3/2}$ orbital of
$^{12}$C, and the dotted line the $1p^{3/2}$ orbital of
\hspace{0.15cm} $_{\Lambda}^{12}$ \hspace{-0.25cm} B. For this figure we only
employed the FSUGold model.
\label{fig_8}}
\end{figure}

In Fig.~\ref{fig_8} we show the radial momentum space wave functions for the proton and the lambda. For the
proton wave function we only employed the FSUGold model. 
The binding energies for the proton and
lambda for the different orbitals of $^{12}$C and $_{\Lambda}^{12}$ \hspace{-0.25cm} B is shown
in Table~\ref{table_1}.
These binding energies are needed for the calculation of the magnitude of
the three-momentum of the outgoing kaon. See Eq.~(\ref{eq_17}).

\begin{figure}
\includegraphics[height=8cm,angle=0]{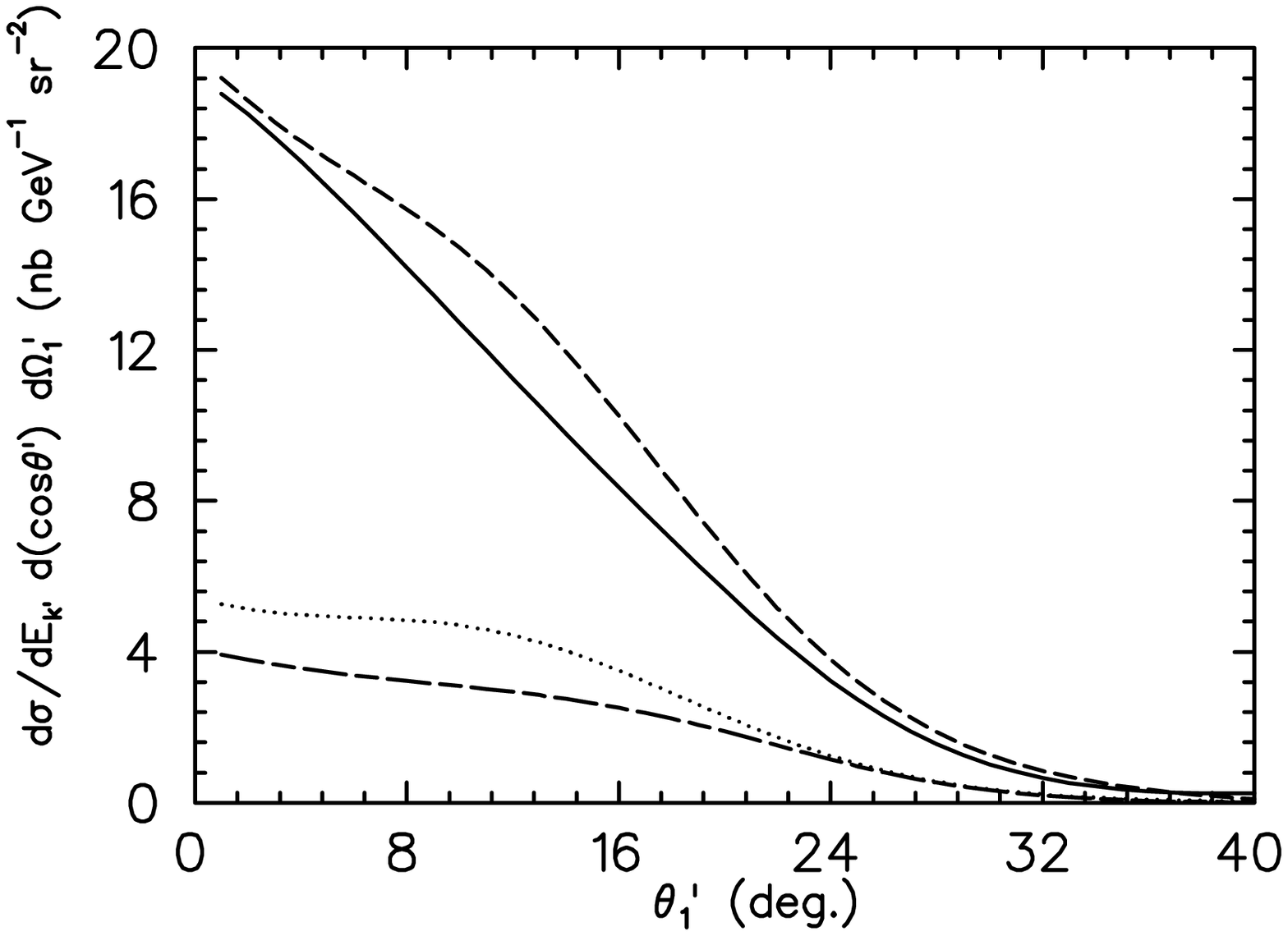}
 \caption{Unpolarized triple differential cross section for the hypernucleus production
process
e + $^{12}$C \, $ \longrightarrow $ \, e + $ K^{+} + {^{12}_{~\Lambda}}$B
as a function of the kaon laboratory scattering angle $\theta_{1}'$.
The incident electron laboratory kinetic energy is $E_{k} \, = \, 3$ GeV, the outgoing
electron laboratory scattering angle is $\theta' \, = \, 5^{\circ}$, the outgoing electron laboratory kinetic energy is
$E_{k'} \, = \, 2$ GeV, and the angle between the
leptonic and hadronic planes is $\phi' \, = \, 0^{\circ}$. The solid line represents the calculation for a proton in
the $1p^{3/2}$ orbital, and a $\Lambda$ in the $1p^{3/2}$ orbital, the dashed line a proton in the $1p^{3/2}$ orbital
and a $\Lambda$ in the $1s^{1/2}$ orbital, the long-dashed--short-dashed line a proton in the $1s^{1/s}$ orbital
and the $\Lambda$ in the $1p^{3/2}$ orbital, and the dotted line a proton in the $1s^{1/2}$ orbital and the $\Lambda$
in the $1s^{1/2}$ orbital.
\label{fig_9}}
\end{figure}

Next we display in Fig.~\ref{fig_9} results for the unpolarized differential cross section (Eq.~(\ref{eq_16b}))
as a function of the kaon scattering angle $\theta_{1} '$, based on the following choice of kinematics:
$\{ E_{k}, \theta', E_{k'}, \phi' \, \} \, = \, \{ 3 \, \mbox{GeV}, 5^{\circ}, 2 \, \mbox{GeV}, 0^{\circ} \}$.
For the hypernuclear production process given in Eq.~(\ref{eq_88}) there are four particle-hole transitions which may
be studied within our simplified model-dependent form for ${\cal W}^{\mu \nu}$ 
(see Eqs.~(\ref{eq_55}),(\ref{eq_56}), (\ref{eq_57}) and (\ref{eq_61})). The cross section has the same behavior for
all the possible transitions namely, large for small angles and a smooth fall-off to zero with increasing angle. The two
upper cross sections (solid and dashed lines) correspond to the probe interacting with a $1p^{3/2}$ valence proton. Of
these, the cross section is highest for a $\Lambda$ in the $1s^{1/2}$ shell. The two lower cross sections (dotted and
long-dashed--short-dashed lines) are for a proton in the $1s^{1/2}$ shell. Again, the $1s^{1/2}$ $\Lambda$ yields a 
higher cross section. 

\begin{figure}
\includegraphics[height=10cm,angle=0]{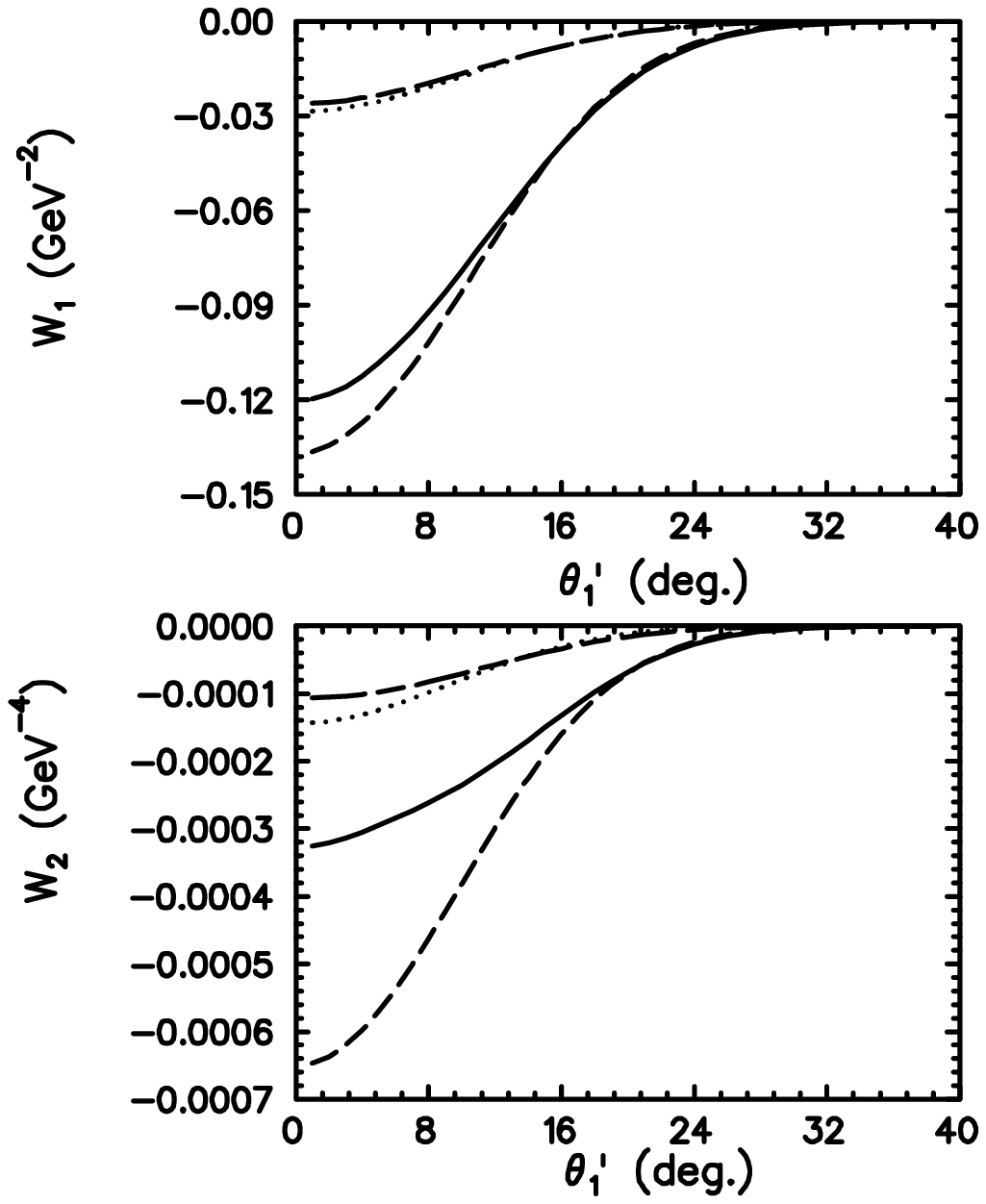}
 \caption{Model-dependent structure functions $W_{1}$ and $W_{2}$ as function of the kaon scattering angle
$\theta_{1}'$.
The solid line represents the calculation for a proton in
the $1p^{3/2}$ orbital, and a $\Lambda$ in the $1p^{3/2}$ orbital, the dashed line a proton in the $1p^{3/2}$ orbital
and a $\Lambda$ in the $1s^{1/2}$ orbital, the long-dashed--short-dashed line a proton in the $1s^{1/s}$ orbital
and the $\Lambda$ in the $1p^{3/2}$ orbital, and the dotted line a proton in the $1s^{1/2}$ orbital and the $\Lambda$
in the $1s^{1/2}$ orbital.
\label{fig_10}}
\end{figure}

\begin{figure}
\includegraphics[height=10cm,angle=0]{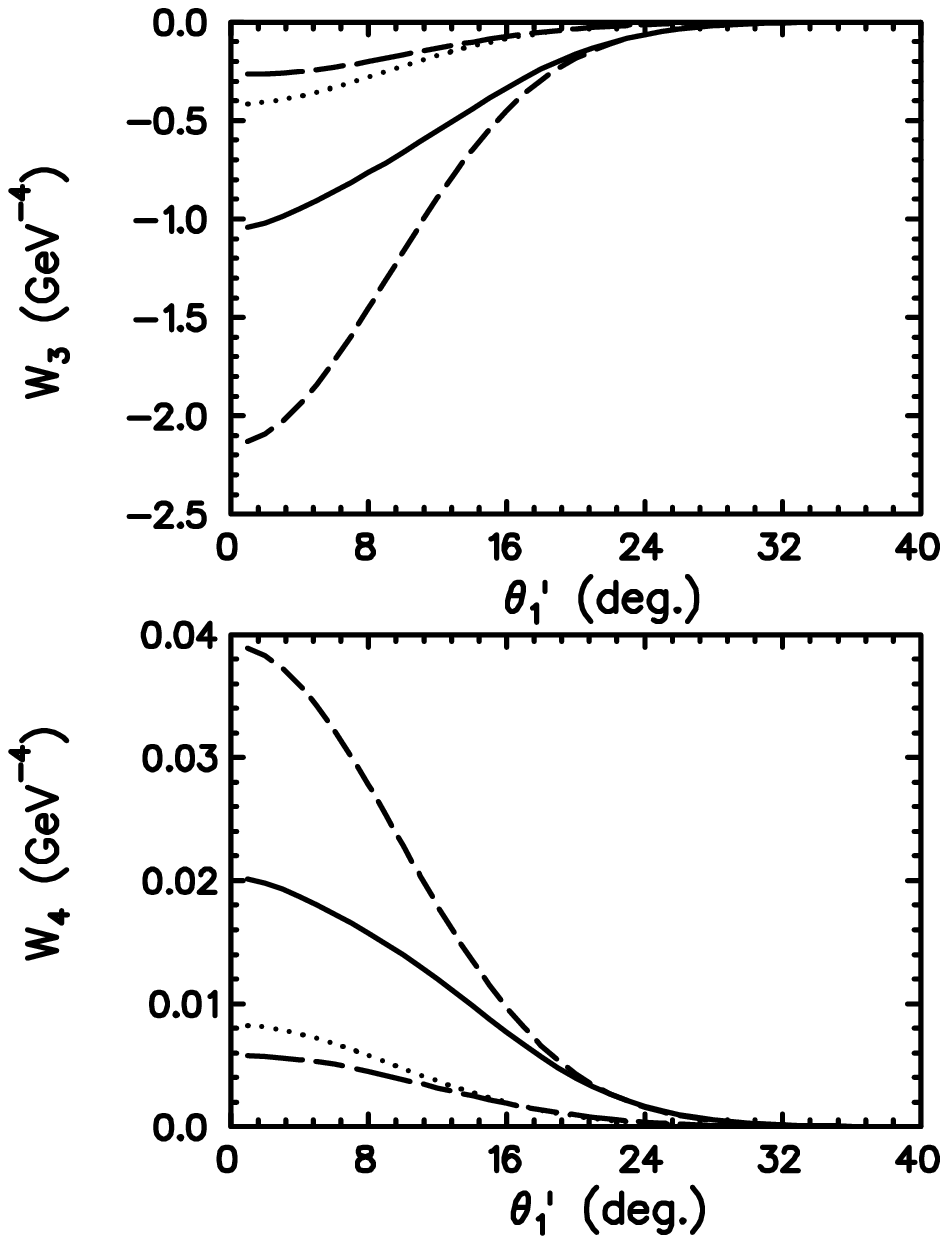}
 \caption{Model-dependent structure functions $W_{3}$ and $W_{4}$ as function of the kaon scattering angle
$\theta_{1}'$.
The solid line represents the calculation for a proton in
the $1p^{3/2}$ orbital, and a $\Lambda$ in the $1p^{3/2}$ orbital, the dashed line a proton in the $1p^{3/2}$ orbital
and a $\Lambda$ in the $1s^{1/2}$ orbital, the long-dashed--short-dashed line a proton in the $1s^{1/s}$ orbital
and the $\Lambda$ in the $1p^{3/2}$ orbital, and the dotted line a proton in the $1s^{1/2}$ orbital and the $\Lambda$
in the $1s^{1/2}$ orbital.
\label{fig_11}}
\end{figure}

\begin{figure}
\includegraphics[height=10cm,angle=0]{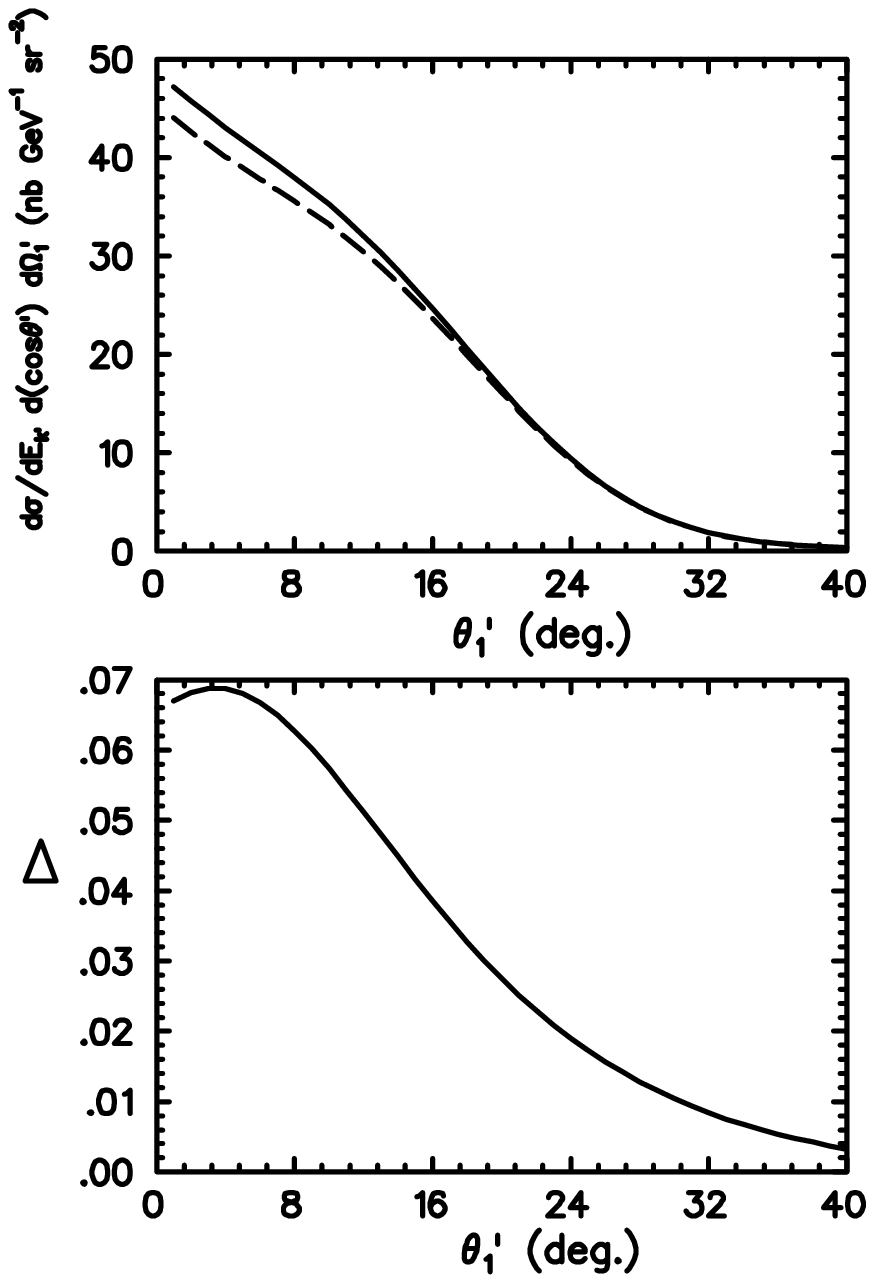}
 \caption{The top graph shows triple differential cross section as a function of the kaon scattering angle
$\theta_{1}'$. The solid line represents the total triple differential cross section, i.e., where a sum over all four
possible transitions has been performed, and the dashed line represents a similar calculation, but where the structure
function $W_{1}$ has been neglected. The bottom graph shows the quantity $\Delta$ defined in Eq.~(\ref{eq_95a}).
\label{fig_12}}
\end{figure}

It is also instructive to plot the model-dependent structure functions $W_{1}$ to $W_{4}$ as a 
function of the kaon scattering angle $\theta_{1}'$. This is shown in Figs.~\ref{fig_10} and \ref{fig_11} for each
of the four possible transitions under consideration. It is clear from these figures that the shape of the cross section
is determined by the structure functions. In Fig.~\ref{fig_12} we plot the total cross section (indicated by the solid
line), i.e., summed over all four possible transitions. The dashed line represents a similar calculation, but where the
structure function $W_{1}$ has been neglected. This graph suggests that $W_{1}$ is negligible for 
$\theta_{1}' \, \ge \, 16^{\circ}$, and makes a very small contribution for angles less than $16^{\circ}$. To quantify
this we show in the bottom figure of Fig.~\ref{fig_12} the quantity $\Delta$ defined as:
\begin{eqnarray}
\label{eq_95a}
 \Delta & = & 
 \left(
 \frac
 {
  {\displaystyle \frac{d \sigma}{d E_{k'} \, d \left( \cos \theta' \, \right) \,d \Omega_{1}'}}
 -
  {\displaystyle \left( \frac{d \sigma}{d E_{k'} \, d \left( \cos \theta' \, \right) \,d \Omega_{1}'} 
\right)_
 {\mbox{{\small no }} \, W_{1}}}}
{\displaystyle \frac{d \sigma}{d E_{k'} \, d \left( \cos \theta' \, \right) \,d \Omega_{1}'}} \right) \times 100 \%
\end{eqnarray}
as a function of $\theta_{1}'$. This extremely small difference illustrates that $W_{1}$ is truly negligible over a wide
angular range. The unpolarized triple differential cross section is therefore essentially just a function of three
structure functions, namely $W_{2}$, $W_{3}$ and $W_{4}$. This could, in principle, allow a Rosenbluth-type analysis,
similar to electron-proton scattering, to be performed for hypernuclei electromagnetic production to disentangle the
various structure functions. 

\begin{figure}
\includegraphics[height=8cm,angle=0]{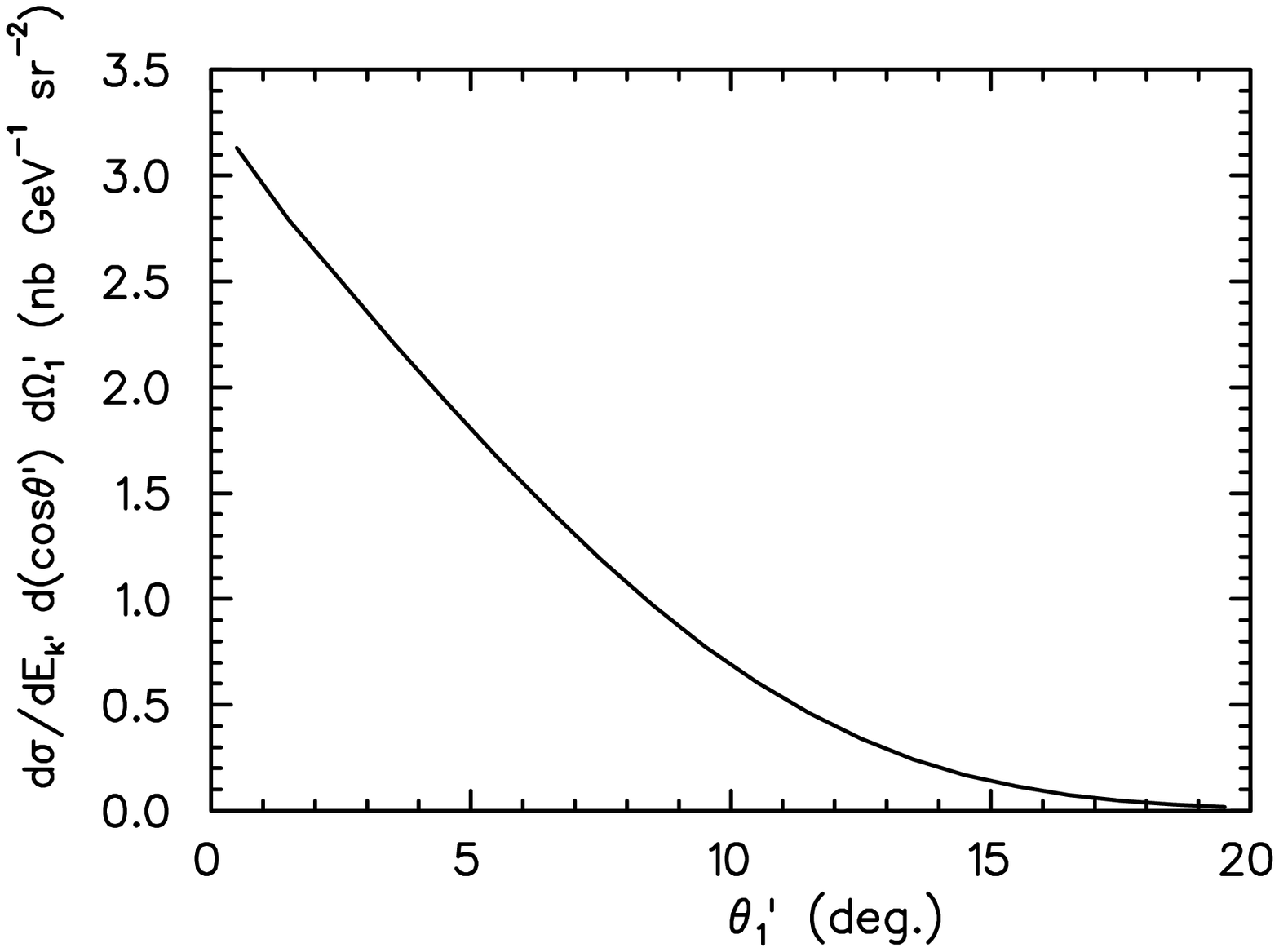}
 \caption{Unpolarized triple differential cross section for the hypernucleus production
process
e + $^{12}$C \, $ \longrightarrow $ \, e + $ K^{+} + {^{12}_{~\Lambda}}$B
as a function of the kaon laboratory scattering angle $\theta_{1}'$.
The incident electron laboratory kinetic energy is $E_{k} \, = \, 1.8$ GeV, the outgoing
electron laboratory scattering angle is $\theta' \, = \, 1^{\circ}$, the outgoing electron laboratory kinetic energy is
$E_{k'} \, = \, 0.5$ GeV, and the angle between the
leptonic and hadronic planes is $\phi' \, = \, 0^{\circ}$. The solid line represents the total cross section, i.e., where
we have summed over all four possible transitions within our simplified particle-hole model for the transition
matrix element.
\label{fig_13}}
\end{figure}

Finally, we show in Fig.~\ref{fig_13} the unpolarized triple differential cross section 
as a function of the kaon scattering angle $\theta_{1}'$, for the kinematical set: 
$\{ E_{k}, \theta', E_{k'}, \phi' \, \} \, = \, \{ 1.8 \, \mbox{GeV}, 1^{\circ}, 0.5 \, \mbox{GeV}, 0^{\circ} \}$.
The calculation shown is for the total cross section, i.e., we have summed over all four
possible transitions allowed within our simplified particle-hole model.
As before the cross section is high for small values of the kaon scattering angle, with
a smooth fall-off to zero as the angle increases.
\begin{table}
\renewcommand{\arraystretch}{1.5}
\caption{\label{table_1} Binding energies for the proton and lambda for the different
orbitals of $^{12}$C and $_{\Lambda}^{12}$ \hspace{-0.25cm} B. The four numbers in square
brackets for the proton orbitals are the predictions for the binding energy (in MeV) of the 
different models in the order QHDI, QHDII, NL3 and FSUGold.}
\begin{ruledtabular}
\begin{tabular}{cc}
 proton $^{12}$C & lambda $_{\Lambda}^{12}$B \\
\hline\\[-2ex]
 $1s^{1/2}$ \hspace{1cm} [42.97, 38.99, 49.70, 39.19] & $1s^{1/2}$ \hspace{1cm} [12.31]\\
 $1p^{3/2}$ \hspace{1cm} [16.17, 12.86, 16.04, 13.67] & $1p^{3/2}$ \hspace{1cm} [1.11]\\
\end{tabular}
\end{ruledtabular}
\end{table}

\section{\label{section_summary}Summary}

In this work a momentum space formalism was developed for the electromagnetic production of hypernuclei. The basic
philosophy is to write the cross section as a contraction of a leptonic tensor and a hadronic tensor. The leptonic
tensor is dependent on the helicity of the incoming and outgoing electron beams. We can therefore calculate
fully, partially or unpolarized triple differential cross sections. The hadronic tensor is written in terms of five
nuclear structure functions. The merit of writing the cross section in this way is that it could in principle allow
a Rosenbluth-type separation to investigate the nature of the hadronic tensor. In this work we have not explored this
avenue, but instead calculated the hadronic tensor based on the following model: it is assumed that the virtual photon
interacts with only one bound nucleon in the nucleus, and that the elementary operator is left unchanged when it is
embedded in the nuclear medium, i.e., we invoke the impulse approximation.
There are two critical ingredients when calculating
hypernuclei cross sections. Firstly, the model that is used to describe the elementary operator,
and secondly, the model that is adopted to describe the nuclear structure. 
To calculate the elementary current operator we first expand it in terms of a set of six invariant
amplitudes. These invariant amplitudes are calculated by writing down the Born diagrams, as well as
the $s$-, $t$-, and $u$-channel Feynman diagrams. The following nucleon and meson resonances are included
namely, $S_{11}(1650), P_{11}(1710), P_{13}(1720), D_{13}(1895),
K^{*}(892)$ and $K_{1}(1270)$.
In our model nuclear structure, which enter exclusively in terms of the momentum distribution of
the bound nucleon and bound hyperon, are calculated within the framework of relativistic mean-field
theory. We considered three models namely the linear Walecka model as well as the NL3 and FSUGold models.
On the wave function level there is not a very big difference between the models and therefore we
employed the FSUGold model for all our cross section calculations.
As a first application of our formalism we calculated the unpolarized
triple differential cross section for hypernuclear electroproduction
from $^{12}$C. A simple model for the transition matrix element was
adopted which only included particle-hole transitions. The calculations indicate that the cross section
is high for small values of the kaon scattering angle, and falls off smoothly to zero with increasing
angle. The cross section for the four possible transitions has a specific structure. 
The cross section for transitions from the $1p^{3/2}$ proton shell is
larger than transitions from the $1s^{1/2}$ proton shell. In turn, for a specific proton shell, the cross section
for the $1s^{1/2}$ lambda shell is higher. The individual model-dependent structure functions were also calculated.
The results indicate that the shape of the structure functions is remarkably similar to the shape of the cross section.
In addition, it is found that the $W_{1}$ structure function is negligible over a wide angular range of
the kaon scattering angle. This indicates that the unpolarized triple differential cross section is essentially just
determined by three structure functions. This could, in principle, allow a Rosenbluth-type analysis to be performed for
for hypernuclear electromagnetic production. 
There are many other questions which may addressed, such as the role of resonances in hypernuclear
production compared to the free process or the role of spin observables as an additional tool to study how sensitive the
hypernuclear cross section is to the elementary operator. Our formalism also allows the study of possible medium effects
on the resonances. Further improvements also need to be made to the calculation of the transition matrix element which
in this work was based on a simple particle-hole model. However, we have established a model-independent form of the
unpolarized and polarized cross sections in terms of nuclear structure functions. Improvements to the calculation of
the transition matrix element will therefore only impact on the hadronic tensor.

\begin{acknowledgments}
This material is based upon work supported by the
National Research Foundation under Grant numbers GUN 2048567
(B.I.S.v.d.V). GUN 2054166 (G.C.H), GUN 2067864 (H-F.L) and 
GUN 2067863 (H.L.Y). The work of T.M. was supported by the
University of Indonesia.
\end{acknowledgments}

\appendix
\section{Kinematics for electromagnetic hypernuclei production}
\label{section_appendix1}

\begin{figure}
\includegraphics[height=8cm,angle=0]{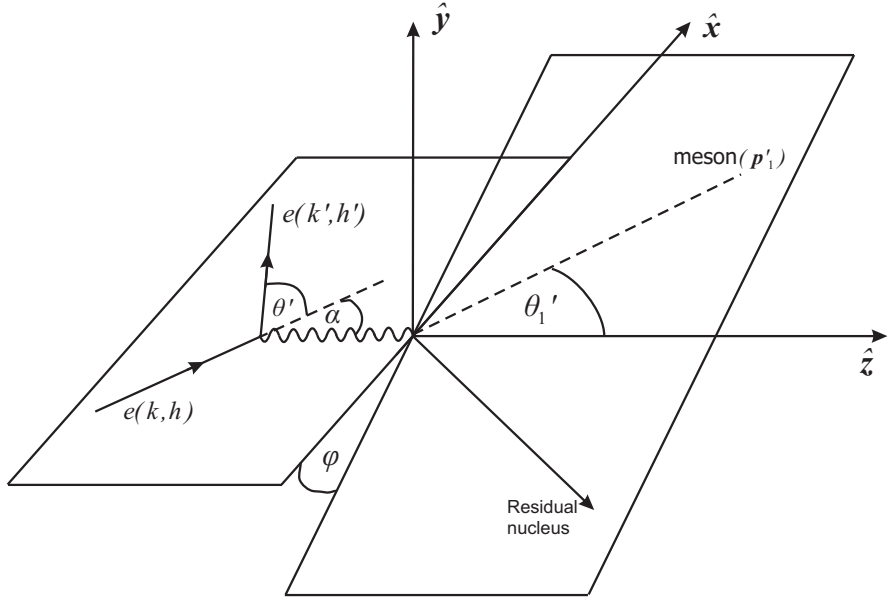}
 \caption{Leptonic and hadronic planes for hypernuclei electroproduction.
\label{fig_2}}
\end{figure}

As is shown in Sec.~\ref{section_cross_section} the differential cross section
is given by
\begin{eqnarray}
\label{eq_2a}
 d \sigma & = & \displaystyle \frac{1}{| {\bf v_{1}} - {\bf v_{2}} \, |} \,
\frac{d^{3} {\bf k}'}{(2 \pi)^{3}} \, \frac{d^{3} {\bf p_{1}'}}{2 E_{p_{1}'} (2 \pi)^{3}} \,
\frac{d^{3} {\bf P}'}{(2 \pi)^{3}} \, (2 \pi)^{4} \, \delta(k + P - k' - p_{1}' - P' \, ) \,
|{\cal M}|^{2}.
\end{eqnarray}
We can simplify Eq.~(\ref{eq_2a}) by employing the spatial part of the four-dimensional Dirac
delta function to do the integral over the three-momentum of the residual hypernucleus which leads to
\begin{eqnarray}
\label{eq_4}
 d \sigma & = & \displaystyle \frac{\delta(E_{k} + M_{A} - E_{k'} - E_{p_{1}'} - E_{P'})}
 {2 (2 \pi)^{5} \, E_{p_{1}'}} \, d^{3} {\bf k'} \, d^{3} {\bf p_{1}'} \, |{\cal M}|^{2}.
\end{eqnarray}
In order to derive a triple differential cross section which may be compared to experiment,
it is necessary to derive expressions that fully specify the four-vectors of the incoming
and outgoing electrons, as well as the outgoing meson. The kinematical set-up for 
hypernuclei production is shown in Fig.~\ref{fig_2}. The direction of the virtual photon
three-momentum defines the $\hat{{\bf z}}$-axis, i.e.,
\begin{eqnarray}
\label{eq_5}
 \hat{{\bf z}} & = & \displaystyle \frac{{\bf q}}{|{\bf q}|}.
\end{eqnarray}
The unit vectors $\hat{{\bf x}}$ and $\hat{{\bf y}}$ define the leptonic plane in 
Fig.~\ref{fig_2}. 

The
right-handed coordinate system is completed by defining
\begin{eqnarray}
\label{eq_6}
 \hat{{\bf y}} & = & \hat{{\bf z}} \times \hat{{\bf x}}.
\end{eqnarray}
In the leptonic plane the electron scattering angle is $\theta'$, and the direction of the incident
electron with respect to the $\hat{{\bf z}}$ is denoted by the angle $\alpha$. In the hadronic plane the
meson scattering angle is denoted by $\theta_{1}'$. The hadronic plane makes an angle $\phi$ with respect to the
leptonic plane. The three-momentum of the incoming electron ${\bf k}$ (defined with respect to the coordinate system in
Fig.~\ref{fig_2}) is given by, for massless electrons,
\begin{eqnarray}
\label{eq_7}
 {\bf k} & = & \left( E_{k} \sin \alpha \right) \, {\bf \hat{x}} + 
 \left( E_{k} \cos \alpha \right) \, {\bf \hat{z}}.
\end{eqnarray}
The energy transfer to the nucleus is given by
\begin{eqnarray}
\label{eq_8}
 q_{0} & = & E_{k} - E_{k'}.
\end{eqnarray}
The three-momentum of the outgoing electron ${\bf k'}$ is given by
\begin{eqnarray}
\label{eq_9}
 {\bf k'} & = & \left[ E_{k'} \sin (\alpha + \theta') \right] {\bf \hat{x}} +
 \left[ E_{k'} \cos (\alpha + \theta') \right] {\bf \hat{z}}.
\end{eqnarray}
Since the virtual photon direction defines the ${\bf \hat{z}}$-axis, it follows that
${\bf q} \, = \, (0,0,|{\bf q}|)$, and therefore we can determine the angle $\alpha$ in 
Fig.~\ref{fig_2} by demanding that $q_{x} \, = \, k_{x} - k_{x}'$ should be zero.
This yields the following equation for the angle $\alpha$: 
\begin{eqnarray}
\label{eq_10}
 \sin^{2} \alpha & = & \displaystyle \frac{E_{k'}^{2} \, \sin^{2} \theta'}
 {E_{k}^{2} + E_{k'}^{2} - 2 E_{k} E_{k'} \, \cos \theta'}.
\end{eqnarray}
Geometric arguments show that the laboratory three-momentum of the outgoing meson
is given by
\begin{eqnarray}
\label{eq_11}
 {\bf p_{1}'} & = & \sqrt{E_{p_{1}'} - M_{K}^{2}} \,
 \left[ \, \left( \sin \theta_{1}' \cos \phi' \right) \, {\bf \hat{x}} + 
 \left( \sin \theta_{1}' \sin \phi' \right) \, {\bf \hat{y}} +
 \left( \cos \theta_{1}' \right) \, {\bf \hat{z}} \, \right].
\end{eqnarray}
To proceed any further in specifying the kinematics, we need to determine the
energy of the outgoing meson $E_{p_{1}'}$. This is done by using the Dirac
delta function in Eq. (\ref{eq_4}). The total energy of the residual hypernucleus is
given by
\begin{eqnarray}
\label{eq_12}
 E_{P'} & = & \sqrt{{\bf P'} + _{\Lambda} \hspace{-0.15cm} M_{A}} \, = \,
 \sqrt{\left( {\bf q} - {\bf p_{1}'} \, \right)^{2} + _{\Lambda} \hspace{-0.15cm} M_{A}}
\\
\label{eq_13}
 & = &
 \sqrt{\left( {\bf q} - {\bf p_{1}'} \, \right)^{2} + 
 \left[ M_{A} - \left( M_{p} - E_{B}^{(N)} \right) + 
 \left( M_{Y} - E_{B}^{(Y)} \right) \, \right]^{2}}
\end{eqnarray}
where $E_{B}^{(N)}$ and $E_{B}^{(Y)}$ are the bound state energies for the nucleon and
hyperon orbitals, respectively. These two quantities will be obtained from a relativistic
mean-field model of the nuclear structure (see Sec.~\ref{section_nuclear_structure}).
The quantity $E_{P'}$ can be written as a function of the outgoing meson energy
$E_{p_{1}'}$. In addition, using the relations
\begin{eqnarray}
\label{eq_14}
 d^{3} {\bf k'} & = & 2 \pi E_{k'}^{2} \, d E_{k'} \, d \left( \cos \theta' \, \right),
\end{eqnarray}
\begin{eqnarray}
\label{eq_15}
 d^{3} {\bf p_{1}'} & = & \sqrt{E_{p_{1}'}^{2} - M_{K}^{2}} \, E_{p_{1}'} \, 
 d E_{p_{1}'} \, d \Omega_{1}'
\end{eqnarray}
the triple differential cross section for the electromagnetic production of 
hypernuclei can be written as
\begin{eqnarray}
\label{eq_16}
 \displaystyle \frac{d \sigma}{d E_{k'} \, d \left( \cos \theta' \, \right) \,
 d \Omega_{1}'}
 & = &
 \displaystyle \frac{E_{k'}^{2} \, \sqrt{E_{p_{1}'}^{2} - M_{K}^{2}}}
 {2 \left( 2 \pi \right)^{4} |f' (E_{p_{1}'}) \, | } \, 
 |{\cal M} |^{2}
\end{eqnarray}
where the energy of the outgoing meson $E_{p_{1}'}$ is obtained by finding the roots of
the equation
\begin{eqnarray}
\nonumber
 f (E_{p_{1}'}) & = & E_{k} + M_{A} - E_{k'} - E_{p_{1}'} -
 \left[ {\bf q}^{2} + E_{p_{1}'}^{2} - M_{K}^{2} - 2 |{\bf q}| \,
 \sqrt{E_{p_{1}'}^{2} - M_{K}^{2}} \, \cos \theta_{1}' + \right. 
\\
\label{eq_17}
 & & 
\left. \left( M_{A} - M_{p} + M_{Y} + E_{B}^{(N)} - E_{B}^{(Y)} \, \right)^{2} \, 
\right]^{\frac{1}{2}}
\end{eqnarray}
where
\begin{eqnarray}
\label{eq_18}
 |{\bf q}| & = & \left[ E_{k}^{2} + E_{k'}^{2} - 2 E_{k} \, E_{k'} \,
 \cos \theta' \, \right]^{\frac{1}{2}}.
\end{eqnarray}
We can see that the leptonic four-vectors, as well as the outgoing meson four-vector
are fully determined if we specify the following kinematical quantities
$\left[ E_{k}, E_{k'}, \theta', \Omega_{1}' \equiv \left( \theta_{1}', \phi' \, \right) \, \right]$. 

\section{Leptonic and hadronic tensors}
\label{section_appendix2}

The transition matrix element ${\cal M}$ for the electromagnetic production of hypernuclei
may be defined as
\begin{eqnarray}
\label{eq_19}
 {\cal M} & = & \left[ \overline{U} ({\bf k'}, h' \, ) \gamma_{\mu} \, 
 U ({\bf k},h) \, \right] \, \left( \frac{e^{2}}{q^{2}} \right) \,
 \langle \, p_{1}'; \, _{\Lambda} \hspace{-0.10cm} \Psi (P' \, ) \, | \hat{J}^{\mu} (q) \, | 
 \Psi(P) \, \rangle
\end{eqnarray}
with $e^{2}/4 \pi \, = \, 1/137$. 
In Eq. (\ref{eq_19}) $\hat{J}^{\mu}$ is the nuclear current operator.
Since we are neglecting the electron mass with respect to the
total energy, we employ the helicity representation \cite{VanderVentel_PRC69_035501_2004}
of the plane wave Dirac spinor
\begin{eqnarray}
\label{eq_20}
 U ({\bf k},h) 
 & = &
 \displaystyle \frac{1}{\sqrt{2}} \, \left(
  \begin{array}{c}
   \phi_{h} ({\bf \hat{k}})
   \\[0.5cm]
   h \phi_{h} ({\bf \hat{k}})
  \end{array}
 \right)
\end{eqnarray}
with
\begin{eqnarray}
\label{eq_21}
 \phi_{h'} ({\bf \hat{k}'})
 & = &
 \left(
  \begin{array}{c}
   \delta_{h', 1} \, \cos \displaystyle \frac{\theta'}{2} -
   \delta_{h', -1} \, e^{-i \phi'} \, \sin \displaystyle \frac{\theta'}{2}
   \\[0.25cm]
   \delta_{h', 1} \, e^{i \phi'} \, \sin \displaystyle \frac{\theta'}{2}
   + \delta_{h', -1} \, \cos \displaystyle \frac{\theta'}{2}
  \end{array}
 \right)
\end{eqnarray}
where the unit vector ${\bf \hat{k}}$ is specified by the polar and azimuthal angles, 
$\theta$ and $\phi$, respectively. This spinor is non-covariantly normalized to
$U^{\dagger} U \, = \, 1$, and corresponds to the normalization adopted for the
bound state spinors. See Sec. \ref{section_Wmunu_model}.
It follows that $|{\cal M}|^{2}$ can be written as a contraction between the 
leptonic and hadronic tensors, i.e.,
\begin{eqnarray}
\label{eq_22}
 |{\cal M}|^{2} 
 & = &
 \left( \frac{e^{2}}{q^{2}} \right)^{2} \, \ell_{\mu \nu} \, {\cal W}^{\mu \nu}
\end{eqnarray}
where
\begin{eqnarray}
\label{eq_23}
 \ell_{\mu \nu}
 & = & 
 \left[ \, \overline{U} ({\bf k'}, h' \, ) \gamma_{\mu} \, U ({\bf k},h) \, \right] \,
 \left[ \, \overline{U} ({\bf k'}, h' \, ) \gamma_{\nu} \, U ({\bf k},h) \, \right]^{*}
\end{eqnarray}
and
\begin{eqnarray}
\label{eq_24}
 {\cal W}^{\mu \nu}
 & = &
 \langle \, p_{1}'; \, _{\Lambda} \hspace{-0.10cm} \Psi (P' \, ) \, | \hat{J}^{\mu} (q) \, | 
 \Psi(P) \, \rangle \,
 \langle \, p_{1}'; \, _{\Lambda} \hspace{-0.10cm} \Psi (P' \, ) \, | \hat{J}^{\nu} (q) \, | 
 \Psi(P) \, \rangle^{*}.
\end{eqnarray}
We will now study each one of these tensors in detail. The leptonic tensor may be written
as a trace over Dirac matrices. This is done by using the identity
\begin{eqnarray}
\label{eq_25}
 U ({\bf k},h) \, \overline{U} ({\bf k'}, h' \, )
 & = &
 \displaystyle \frac{\rlap/k}{4 E_{k}} \, \left( 1 - h \gamma^{5} \, \right).
\end{eqnarray}
It follows from Eq. (\ref{eq_25}) that the leptonic tensors will in general be dependent
on the helicity of the incoming and outgoing electrons. We may therefore distinguish
four different cases for the leptonic tensor. In the first case both the incident and
outgoing electron beams are unpolarized.
In this case we define
\begin{eqnarray}
\label{eq_26}
 \ell^{(0)}_{\mu \nu}
 & = &
 \mbox{Tr} \, \left[ \gamma_{\mu} \; \left( \sum_{h \, = \, \pm 1} \, 
 U ({\bf k},h) \, \overline{U} ({\bf k},h) \, \right) \, \gamma_{\nu} \, 
 \left( \sum_{h' \, = \, \pm 1} \, U ({\bf k'},h') \, \overline{U} ({\bf k'},h' \, ) 
 \, \right) \, \right]
\\
\label{eq_27}
 & = &
 \displaystyle \frac{1}{E_{k} \, E_{k'}} \,
 \left( k_{\mu} \, k_{\nu}' + k_{\mu}' k_{\nu} - k \cdot k' \, g_{\mu \nu} \right).
\end{eqnarray}
Note that $\ell^{(0)}_{\mu \nu}$ is completely symmetric in $\mu$ and $\nu$.
In the second case the incident electron beam is polarized and the outgoing electron beam is unpolarized.
In this case we define
\begin{eqnarray}
\label{eq_28}
 \ell^{(1)}_{\mu \nu} (k,h; k' \, )
 & = &
 \mbox{Tr} \left[ \gamma_{\mu} \, \left( U ({\bf k},h) \, \overline{U} ({\bf k},h) \, \right) \,
 \gamma_{\nu} \, \left( \sum_{h' \, = \, \pm 1} \, U ({\bf k'},h') \, \overline{U} ({\bf k'},h' \, ) \right) \, \right]
\\
\label{eq_29}
 & = &
 \displaystyle \frac{1}{2} \, \ell^{(0)}_{\mu \nu} - 
 \displaystyle \frac{ih}{2 E_{k} E_{k'}} \, k^{\alpha} \, k'^{\beta} \,
 \epsilon_{\mu \nu \alpha \beta}
\end{eqnarray}
where we adopt the convention $\epsilon^{0123} \, = \, +1$ for the Levi-Civita tensor.
Note that the lepton tensor now contains an anti-symmetric term due to the polarization of the
incoming electron beam. If the incident beam is unpolarized and the outgoing beam is polarized
then we define
\begin{eqnarray}
\label{eq_30}
 \ell^{(2)}_{\mu \nu} (k; k',h' \, )
 & = &
 \mbox{Tr} \left[ \gamma_{\mu} \, \left( \sum_{h' \, = \, \pm} \, U ({\bf k},h) \, \overline{U} 
 ({\bf k},h) \right) \, \gamma_{\nu} \, \left( U ({\bf k'},h') \, \overline{U} ({\bf k'},h') \right) \, 
 \right]
\\
 \label{eq_31}
 & = &
 \ell^{(1)}_{\nu \mu} (k',h'; k).
\end{eqnarray}
In the final case the incoming and outgoing beams are polarized. Now
\begin{eqnarray}
\label{eq_32}
 \ell_{\mu \nu} (k,h; k',h' \,)
 & = &
 \mbox{Tr} \, \left[ \gamma_{\mu} \, \left( U({\bf k},h) \, \overline{U} ({\bf k},h) \, \right) 
 \, \gamma_{\nu} \, \left( U({\bf k'},h') \, \overline{U} ({\bf k'},h' \, ) \, \right) \, \right]
\\
\label{eq_33}
 & = &
 \displaystyle \frac{1 + h h'}{4} \, \ell^{(0)}_{\mu \nu} +
 \displaystyle \frac{i(h + h' \, )}{4 E_{k} E_{k'}} \, k^{\alpha} \, k'^{\beta} \,
 \epsilon_{\mu \nu \alpha \beta}. 
\end{eqnarray}
Next we turn our attention to the hadronic tensor ${\cal W}^{\mu \nu}$. The definition
in Eq. (\ref{eq_24}) shows that this is an extremely complicated object since
it contains exact many-body wave functions. However, ${\cal W}^{\mu \nu}$ can only be
a function of the three independent four-momenta, namely $q^{\mu}$, $P^{\mu}$ and
$p_{1}'^{\mu}$. Note that four-momentum conservation fixes
\begin{eqnarray}
\label{eq_34}
 P' & = & q + P - p_{1}'.
\end{eqnarray}
We can therefore expand ${\cal W}^{\mu \nu}$ in terms of a basis constructed from
$\{ g^{\mu \nu}, q^{\mu}, P^{\mu}, p_{1}'^{\mu} \}$. This is similar to
the approach in Ref.~\cite{VanderVentel_PRC69_035501_2004} with the exception that a
parity conserving electromagnetic current forbids the presence of any terms linear
in the Levi-Civita tensor. The expansion for ${\cal W}^{\mu \nu}$ then assumes the
form
\begin{eqnarray}
\nonumber
 {\cal W}^{\mu \nu}
 & = &
 W_{1}' \, g^{\mu \nu} + W_{2}' \, q^{\mu} q^{\nu} + W_{3}' \, P^{\mu} P^{\nu} +
 W_{4}' \, p_{1}'^{\mu} p_{1}'^{\nu} + 
 W_{5}' \, \left( P^{\mu} q^{\nu} + q^{\mu} P^{\nu} \, \right) +
\\
\nonumber
 & &
 W_{6}' \, \left( p_{1}'^{\mu} q^{\nu} + q^{\mu} p_{1}'^{\nu} \, \right) +
 W_{7}' \, \left( p_{1}'^{\mu} P^{\nu} + P^{\mu} p_{1}'^{\nu} \, \right) +
 W_{8}' \, \left( P^{\mu} q^{\nu} - q^{\mu} P^{\nu} \, \right) +
\\
\label{eq_35}
 & &
 W_{9}' \, \left( p_{1}'^{\mu} q^{\nu} - q^{\mu} p_{1}'^{\nu} \, \right) +
 W_{10}' \, \left( p_{1}'^{\mu} P^{\nu} - P^{\mu} p_{1}'^{\nu} \, \right).
\end{eqnarray}
At this point ${\cal W}^{\mu \nu}$ contains ten independent nuclear structure functions.
However, the imposition of electromagnetic current conservation, i.e.,
\begin{eqnarray}
\label{eq_36}
 q_{\mu} {\cal W}^{\mu \nu} \, = \, q_{\nu} {\cal W}^{\mu \nu} & = & 0
\end{eqnarray}
reduces the hadronic tensor to the following form \cite{Picklesimer_PRC32_1312_1985}
\begin{eqnarray}
\nonumber
 {\cal W}^{\mu \nu}
 & = &
 W_{1} \, G^{\mu \nu} + W_{2} \, A^{\mu} A^{\nu} + W_{3} \, B^{\mu} B^{\nu} +
 W_{4} \, \left( A^{\mu} B^{\nu} + B^{\mu} A^{\nu} \, \right) +
\\
\label{eq_37}
 & &
 W_{5} \, \left( A^{\mu} B^{\nu} - B^{\mu} A^{\nu} \, \right)
\end{eqnarray}
where
\begin{eqnarray}
\label{eq_38}
 G^{\mu \nu} & = & g^{\mu \nu} - \displaystyle \frac{q^{\mu} q^{\nu}}{q^{2}}
\\
\label{eq_39}
 A^{\mu} & = & P^{\mu} - \displaystyle \frac{P \cdot q}{q^{2}} \, q^{\mu}
\\
\label{eq_40}
 B^{\mu} & = & p_{1}'^{\mu} - \displaystyle \frac{p_{1}' \cdot q}{q^{2}} \, q^{\mu}.
\end{eqnarray}
Note that
\begin{eqnarray}
\label{eq_41}
 A \cdot q \, = \, B \cdot q & = & 0.
\end{eqnarray}
The hadronic tensor consists of four terms which are symmetric with respect to $\mu$
and $\nu$, and the last term which is anti-symmetric with respect to $\mu$ and $\nu$,
hence
\begin{eqnarray}
\label{eq_42}
 {\cal W}^{\mu \nu} & = & {\cal W}_{S}^{\mu \nu} + {\cal W}_{A}^{\mu \nu} \, = \,
 \sum_{i \, = \, 1}^{4} \, W_{i} \, u_{i}^{\mu \nu} + W_{5} \, u_{5}^{\mu \nu}
\end{eqnarray}
where
\begin{eqnarray}
\label{eq_42a}
 \{ u_{1}^{\mu \nu}, u_{2}^{\mu \nu}, u_{3}^{\mu \nu}, u_{4}^{\mu \nu}, u_{5}^{\mu \nu}\}
 & = &
 \{ G^{\mu \nu}, A^{\mu} A^{\nu}, B^{\mu} B^{\nu}, A^{\mu} B^{\nu} + B^{\mu} A^{\nu},
 A^{\mu} B^{\nu} - B^{\mu} A^{\nu} \}.
\end{eqnarray}
Since the contraction of a symmetric and an anti-symmetric tensor is zero, it follows 
immediately that
\begin{eqnarray}
\label{eq_43}
 W_{5} & = & \displaystyle \frac{u_{5,\mu \nu} \, {\cal W}^{\mu \nu}}
 {u_{5, \mu \nu} \, u_{5}^{\mu \nu}}.
\end{eqnarray}
Since the basis $\{ u_{i}^{\mu \nu} \}$ is not orthogonal, we can determine the 
structure functions $W_{i} \, (i \, = \, 1, 2, 3, 4) $ by solving the following set of 
coupled linear equations
\begin{eqnarray}
\label{eq_44}
 \underline{W}_{U} & = & U \, \underline{W}
\end{eqnarray}
where
\begin{eqnarray}
\label{eq_45}
 \left( \underline{W}_{U} \, \right)_{i}
 & = &
 u_{i, \mu \nu} \, {\cal W}^{\mu \nu}, \qquad i \, = \, 1, 2, 3, 4
\end{eqnarray}
and the $4 \times 4$ matrix $U$ is given by
\begin{eqnarray}
\label{eq_46}
 U & = &
 \left(
  \begin{array}{cccc}
   3 & \hspace{0.5cm} A^{2} & \hspace{0.5cm} B^{2} & \hspace{0.5cm} 2 A \cdot B
   \\[0.25cm]
   A^{2} & \hspace{0.5cm} A^{4} & \hspace{0.5cm} \left( A \cdot B \right)^{2} & \hspace{0.5cm}
   2 A^{2} \, \left( A \cdot B \right)
   \\[0.25cm]
   B^{2} & \hspace{0.5cm} \left( A \cdot B \right)^{2} & \hspace{0.5cm} B^{4} & \hspace{0.5cm}
   2 B^{2} \left( A \cdot B \right)
   \\[0.25cm]
   2 A \cdot B & \hspace{0.5cm} 2 A^{2} \, \left( A \cdot B \right) & \hspace{0.5cm} 
   2 B^{2} \, \left( A \cdot B \right)
   & \hspace{0.5cm} 2 A^{2} B^{2} + 2 \left( A \cdot B \right)^{2}
  \end{array}
 \right)
\end{eqnarray}
with
\begin{eqnarray}
\label{eq_47}
 \underline{W} & = & 
 \left(
  \begin{array}{c}
   W_{1}
   \\[0.25cm]
   W_{2}
   \\[0.25cm]
   W_{3}
   \\[0.25cm]
   W_{4}
  \end{array}
 \right).
\end{eqnarray}
Using the general expansion of ${\cal W}^{\mu \nu}$ we can now work out its contraction
with $\ell_{\mu \nu}$. As shown previously, the leptonic tensor can be written in
four different forms, depending on whether the incident and/or outgoing electron 
beams are polarized or not. The four contractions are
\begin{eqnarray}
\nonumber
 |{\cal M}^{(0)} \, |^{2}
 & = &
 \left( \frac{e^{2}}{q^{2}} \right)^{2} \, \ell_{\mu \nu}^{(0)} \, {\cal W}^{\mu \nu}_{S}
\\
\nonumber
 & = &
 \displaystyle \left( \frac{e^{2}}{q^{2}} \right)^{2} \, \frac{1}{E_{k} \, E_{k'}} \,
 \left[ W_{1} \, \left( -3 k \cdot k' + 2 f_{1} (k,k' \, ) \, \right) +
 W_{2} \, \left( -k \cdot k' \, f_{1} (P,P) + \right. \right.
\\
\nonumber
 & &
 \left. 2 f_{1} (k,P) \, f_{1} (k',P) \, \right) + W_{3} \, 
 \left( -k \cdot k' \, f_{1} (p_{1}', p_{1}' \, ) + 2 f_{1} (k, p_{1}') \, f_{1} (k', p_{1}' \, ) \right)  +
\\
\label{eq_48}
 & &
\left. W_{4} \, \left( 2 f_{2} (P,p_{1}' \, ) \, \right) \, 
\right],
\end{eqnarray}
\begin{eqnarray}
 |{\cal M}^{(1)} \, |^{2}
 & = &
 \left( \frac{e^{2}}{q^{2}} \right)^{2} \, 
 \ell_{\mu \nu}^{(1)} (k,h; k' \, ) \, {\cal W}^{\mu \nu}
\\
\label{eq_49}
 & = &
 \displaystyle \left( \frac{e^{2}}{q^{2}} \right)^{2} \,
 \left[ \frac{1}{2} \, \ell_{\mu \nu}^{(0)} \, {\cal W}^{\mu \nu}_{S} -
 \displaystyle \frac{ih}{2 E_{k} \, E_{k'}} \, k^{\alpha} \, k'^{\beta} \,
 \epsilon_{\mu \nu \alpha \beta} \, 
 \left( A^{\mu} B^{\nu} - B^{\mu} A^{\nu} \, \right) \, W_{5} \, \right],
\end{eqnarray}
\begin{eqnarray}
\label{eq_50}
 |{\cal M}^{(2)} \, |^{2}
 & = &
 \left( \frac{e^{2}}{q^{2}} \right)^{2} \, \ell_{\mu \nu}^{(2)} (k; k', h' \, ) \, {\cal W}^{\mu \nu}
 \, = \, 
 \left( \frac{e^{2}}{q^{2}} \right)^{2} \, \ell_{\nu \mu}^{(1)} (k',h'; k) \, {\cal W}^{\mu \nu},
\end{eqnarray}
\begin{eqnarray}
 |{\cal M} \, |^{2}
 & = &
 \left( \frac{e^{2}}{q^{2}} \right)^{2} \, \ell_{\mu \nu} (k, h; k', h' \, ) \, {\cal W}^{\mu \nu}
\\
\label{eq_51}
 & = &
 \left( \frac{e^{2}}{q^{2}} \right)^{2} \,
 \left[ \displaystyle \frac{1 + h h'}{4} \, \ell_{\mu \nu}^{(0)} \, {\cal W}^{\mu \nu}_{S} 
 - \displaystyle \frac{i (h + h' \, )}{4 E_{k} \, E_{k'}} \, k^{\alpha} \, k^{\beta} \, 
 \epsilon_{\mu \nu \alpha \beta} {\cal W}^{\mu \nu}_{A}
 \right]
\end{eqnarray}
where 
\begin{eqnarray}
\label{eq_52}
 f_{1} (x,y) & = & x \cdot y - \displaystyle \frac{x \cdot q \, y \cdot q}{q^{2}} \, = \,
 f_{1} (y,x)
\\[0.25cm]
\label{eq_53}
 f_{2}(x,y) & = & f_{1}(k,x) \, f_{1}(k',y) + f_{1}(k',x) \, f_{1}(k,y) -
 k \cdot k' \, f_{1}(x,y).
\end{eqnarray}
We can now substitute Eqs. (\ref{eq_48}), (\ref{eq_49}), (\ref{eq_50}) or 
(\ref{eq_51}) into Eq. (\ref{eq_16}) to obtain an unpolarized, partially polarized
or fully polarized cross section. The resulting cross section will only depend on 
kinematical quantities (determined by the experimental set-up), and a set of nuclear
structure functions.


\end{document}